\documentclass[a4paper,10pt]{article}
\usepackage{enumerate}
\usepackage{color}
\usepackage[utf8]{inputenc} 
\usepackage[english]{babel}
\usepackage[T1]{fontenc}
\usepackage{graphicx}
\usepackage{amsfonts,amssymb,amsmath,amsthm}
\usepackage{textcomp}

\usepackage[hidelinks]{hyperref}

\usepackage{geometry}
\newcommand{\missingfigurebox}[1]{%
  \fbox{\parbox[c][2.5cm][c]{0.7\linewidth}{\centering\ttfamily\footnotesize
  FIGURE NOT FOUND\\[2pt]\detokenize{#1}}}}
\newcommand{\incfig}[3][]{%
  \IfFileExists{#2}{\includegraphics[#1]{#2}}{%
  \IfFileExists{#3}{\includegraphics[#1]{#3}}{%
  \IfFileExists{Images_draft/#2}{\includegraphics[#1]{Images_draft/#2}}{%
  \IfFileExists{Images_draft/#3}{\includegraphics[#1]{Images_draft/#3}}{%
  \missingfigurebox{#2}}}}}}

\title{ Particle Trajectories Beneath Fully Nonlinear Waves Generated by Horizontal Seabed Motion  }
\author{Jo\~{a}o Vitor P. Poletto$^{1}$, David Andrade$^{2 \footnote{Corresponding author}}$, Marcelo V. Flamarion$^{3}$ and Roberto Ribeiro-Jr$^{1}$}
\date{}

\begin{document}
\maketitle
\begin{center}
{\footnotesize $^1$UFPR/Federal University of Paran\'a,  Departamento de Matem\'atica, Centro Polit\'ecnico, Jardim das Am\'ericas, Caixa Postal 19081, Curitiba, PR, 81531-980, Brazil  \\
}

\vspace{0.3cm}
{\footnotesize 
$^{2}$School of Sciences and Engineering, Universidad del Rosario, 111711, Bogot\'{a}, Colombia.

$^{3}$ {Secci{\' o}n Matem{\' a}ticas, Departamento Acad{\' e}mico} Ciencias, Pontificia Universidad Cat{\' o}lica del Per{\' u}, Av. Universitaria 1801, San Miguel 15088, Lima, Peru}



\end{center}


\begin{abstract}

We investigate fully nonlinear water waves and fluid-particle dynamics generated by the horizontal motion of a seabed obstacle with prescribed time-dependent velocity. The governing equations are the full Euler equations, formulated in a time-dependent conformal domain that simultaneously maps the moving free surface and seabed onto fixed boundaries. The main contribution of this work is a Lagrangian formulation for computing particle trajectories in the resulting genuinely unsteady conformal domain. Previous Lagrangian formulations based on conformal mappings were largely restricted to flows that are steady in a moving frame, so that the particle equations reduce to an autonomous planar system. Landslide-generated flows admit no such frame: the forcing is transient and the conformal map itself is time dependent. We derive a closed-form trajectory system in the canonical domain in which the time dependence of the conformal map is entirely represented by the real and imaginary parts of an analytic function that can be evaluated spectrally from the surface. We apply the formulation to  waves generated by horizontal submarine landslide motion and characterize particle displacements throughout the fluid as functions of the initial particle position and Froude number. Our findings identify distinct regions in which particle motion is predominantly associated with the moving seabed, the generated wave, or the combined action of both. The results reveal a transition in the dominant mechanism driving particle motion: at low Froude numbers, particle displacements are primarily associated with the moving seabed, whereas at high Froude numbers the generated wave becomes increasingly dominant, particularly near the free surface and away from the obstacle path.The numerical predictions are benchmarked against laboratory data available in the literature, showing good agreement and providing a quantitative assessment of the model accuracy. Moreover, laboratory topographies beyond those considered here can be readily incorporated into the numerical framework through a Hermite interpolation procedure.

\end{abstract}

\section{Introduction} \label{Introduction}

This paper is a direct continuation of the work of Poletto et al.~\cite{Poletto:2025}, where we introduced a numerical framework for the full Euler equations over a moving impermeable bottom. The method rests on a double conformal map that flattens the free surface and the time-dependent seabed at once, so that the free-boundary problem becomes a pair of evolution equations posed on a fixed rectangle, integrated spectrally in space and forward in time by means of a Runge--Kutta scheme. In that work, only vertical bottom displacements were considered, as a canonical model for the tsunamigenic uplift caused by a fault rupture. The present paper addresses the complementary generation mechanism namely, a submarine landslide model, as a bottom obstacle moves horizontally with a prescribed, time-dependent velocity. We also expand this framework beyond the free surface: we compute the velocity field throughout the fluid and the Lagrangian dynamics of fluid particles. 

On the physical side, we quantify how far the fully nonlinear solution departs from the linear fully dispersive model that is still the standard tool in this field. Nonlinearity is barely noticeable in the laboratory range explored by Whittaker et al.~\cite{Whittaker:2015}, where $Fr \leq 0.375$ and the linear and nonlinear predictions nearly coincide. As the Froude number approaches and exceeds unity the dynamics deviate from linearity in different and important ways; the linear model underestimates the leading crest, it does not reproduce the pronounced asymmetry between the wave trains radiated in the onshore and offshore directions. A  symmetry breaking occurs, reversing the sign of the obstacle amplitude no longer reflects the solution about the undisturbed level. We show that this departure is controlled almost exclusively by the terminal velocity of the slide, neither the amplitude nor the acceleration of the bottom motion leaves a comparable nonlinear signature and, at $Fr = 1.2$, the leading wave is in fact largest for the smallest acceleration, reversing the trend reported at lower Froude numbers \cite{Whittaker:2017}.

On the numerical side, we propose a numerical scheme that computes particle trajectories under a Dyachenko-type conformal map \cite{Dyachenko:1996,Ruban:2005} when the flow is genuinely unsteady. Combining conformal mapping with the FFT gives the velocity at every point of the fluid at essentially the cost of the surface evolution. This has been exploited repeatedly to study the flow beneath water waves \cite{Chen:2021,Ige:2024,Nachbin_RibeiroJr:2014,Nachbin_RibeiroJr:2017}. In the currently available literature about particle trajectories, the trajectory equations form an {autonomous} system because the flow under consideration is a traveling wave of permanent form, so that passing to a moving frame (where the wave is stationary) removes the time dependence and the particle paths become orbits of a planar autonomous vector field, with stagnation points and phase portraits available as organising structures. 

Landslide-generated flows admit no such frame. The bottom accelerates from rest, travels at constant speed and decelerates again, the wave field is transient, and the conformal map is itself time dependent, so that the canonical coordinates drift with respect to the physical ones. We show that the trajectory system can nonetheless be written in closed form in the canonical domain, with the effect of the moving map collected into two functions $\Gamma$ and $\Lambda$, which are the real and imaginary parts of the single analytic function
\begin{equation*}
w(\xi,\eta,t) \; = \; \frac{X_t + iY_t}{X_{\xi} + iY_{\xi}}.
\end{equation*}
These functions are recovered from their boundary values by solving a Laplace problem on the strip. The outcome is a Lagrangian solver with the same spectral accuracy and essentially the same cost as the Eulerian one, valid for arbitrary unsteady potential flows over moving topography, free of any steadiness assumption. Two exact constraints provide a stringent test of the implementation and are satisfied throughout our simulations: particles released on the free surface remain on it, and particles released on the seabed remain on it, at every instant.

Both advances rely on a third, more technical ingredient. Laboratory obstacles are rarely smooth, and profiles such as semi-ellipses, rectangles or triangles cannot be inserted directly into a spectral solver. We therefore represent the initial bottom profile by a piecewise cubic Hermite interpolating polynomial, which matches both the profile and its slope at the interpolation nodes, and feed the interpolant to the conformal map directly. Thus, the same machinery allows us the handling of arbitrary experimental geometries and vary the shape of the obstacle systematically.

The remainder of this introduction places these results in the context of the existing literature.

Numerous studies have been conducted with the aim of better understanding the effects  of submarine landslides on the generated waves. In particular, we mention the work of Whittaker et al.~\cite{Whittaker:2015}, in which a laboratory setup was employed to investigate the waves generated by a rigid block moving horizontally along the seabed. The authors examined the influence of the Froude number and concluded that it governs the overall wave behavior, a finding that is consistent with the results presented here. Whittaker et al.~analyzed the generated waves within the framework of the linearized Euler equations in order to compare with their experimental data. Indeed, the use of linear or asymptotic models to study waves generated by landslides is widespread in the literature \cite{Fang:2020,Jing:2020,Liu:2020,Lynett:2002,Michele:2022,Sulvianuri:2025}.

Within the framework of non-hydrostatic models, we also mention the works of Firdaus and Behrens \cite{Firdaus:2025_2,Firdaus:2025}, which assume a shallow-water regime. The present model does not rely on this assumption and, as shown in Section~\ref{benchmark}, provides closer agreement with the experimental data.

With regard to more complete models, Whittaker et al., in a later study \cite{Whittaker:2017}, investigated a similar scenario but for higher Froude numbers. In this case, the authors did not employ a linear model to predict the generated waves; instead, they adopted a framework based on the Reynolds-averaged Navier--Stokes (RANS) equations to describe the wave dynamics. In their simulations, the waves were computed using both a $k$--$\varepsilon$ turbulence model and a laminar approach, that is, by solving the RANS equations with molecular viscosity while
neglecting turbulence effects. As shown below, the present model addresses the same problem and produces accurate predictions, the only requirement being that wave breaking does not occur.

Another study addressing waves generated by the horizontal motion of seabed topography is that of Chen et al. \cite{Chen:2024}, in which the authors compared two different numerical models: the Navier--Stokes equations and the Green--Naghdi equations. They investigated the free-surface profile, the pressure at the free surface, and the subsurface velocity field. To this end, the parameters of
the numerical block representing the moving topography were varied, considering both a single translational motion and an oscillatory one. The results showed that the main parameters influencing the generated waves are the disturbance amplitude and its velocity. Moreover, the authors found that both numerical models exhibit good
agreement in most of the simulated cases, indicating that viscous effects on the generated waves are relatively small.

Related studies include the works of Renzi et al. \cite{Renzi:2023} and Jin et al. \cite{Jin:2024}. The former employed a Lagrangian flow model to analyze the free-surface elevation and the subsurface velocity field while varying the terminal velocity of the block, whereas the latter investigated the influence of nonlinear effects induced by a moving circular disturbance on the seabed in supercritical regimes.

Navier--Stokes-based models have been applied successfully to the simulation of waves generated by seabed motions \cite{Chen:2024,Jin:2024,Li:2019}. Their main advantage is the ability to account for viscosity and vorticity, which is absent from formulations based on the full Euler equations and potential flow. The extent to which this matters, for surface waves, is in itself a delicate question \cite{Riquier:2025}. Potential-flow formulations continue to reproduce experimental observations with good accuracy in this setting \cite{Athanassoulis:2019,Chen:2024,Zhao:2026}, while the computational cost of three-dimensional solvers may render them unsuitable for large-scale ocean simulations \cite{Firdaus:2025}. The comparatively low cost of the present approach is what makes the systematic linear--nonlinear comparison reported here feasible, since it requires sweeping the Froude number, the amplitude, the acceleration and the obstacle shape over long propagation times. Throughout this paper we refer to solutions of the full Euler equations as solutions of the nonlinear model.

Within the framework presented here, a related analysis for horizontally moving obstacles was carried out by Wang et al. \cite{Wang:2019}, who investigated waves generated by two landslide bodies moving along the bottom of a channel and compared the linear model, the forced Korteweg--de Vries equation and the full Euler equations while varying the topographic parameters and the landslide velocity. In their simulations, however, the topography moved at constant speed. The present study prescribes a time-dependent velocity, so that the obstacle accelerates from rest towards a terminal velocity and decelerates again, reproducing the forcing actually imposed in the laboratory.

Using the fully nonlinear Euler equations to analyze the effects of landslide motion over a flat bottom, we also cite the work of Zhao et al. \cite{Zhao:2026}. They indicated that the full Euler equations provide an accurate numerical framework for reproducing the experimental results of Whittaker et al.~\cite{Whittaker:2015,Whittaker:2017}, and they investigated the N-wave profiles commonly observed in earthquake-generated tsunamis. Their approach employed a convolution-based smoothing procedure for the semi-elliptical topography, for which the Hermite interpolation described above provides an alternative.

The ability to change the obstacle profile at no additional cost also allows us to ask how much the shape itself matters. Such analyses have previously been conducted within linear and asymptotic theories \cite{Jing:2020,Lo:2021,Sulvianuri:2025} and with a Lagrangian formulation \cite{Renzi:2023}, but, to the best of our knowledge, not within the full Euler equations. Consistently with those studies, we find that the influence of the shape remains small provided the area, the height and the centre of mass of the obstacle are preserved.

The analysis of particle trajectories deserves a separate comment. Experimental investigation of particle motion is considerably more demanding than surface measurement, since it requires resolving the underwater velocity field throughout the evolution of the flow. The numerical framework developed here tracks particles anywhere in the fluid domain and shows how they respond both near and far from the region traversed by the landslide, which is relevant for submarine operations, pollutant dispersion and sediment transport \cite{Berchet:2018,Nachbin_RibeiroJr:2014}. Comparable studies within the full Euler equations were carried out by Nachbin and Ribeiro-Junior for a flat bottom with an underlying current and for an incoming wave interacting with a polygonal topography \cite{Nachbin_RibeiroJr:2014,Nachbin_RibeiroJr:2017}; in both cases the flow was steady in a moving frame, which is the restriction removed here.


The main contributions of this work are the following.
\begin{enumerate}
    \item [i.] A numerical scheme for particle trajectories under a time-dependent conformal map. The scheme applies to non-autonomous flows, for which no change of frame renders the problem steady, and therefore removes the restriction to travelling waves of permanent form that constrains previous conformal-mapping computations of Lagrangian dynamics. The time dependence of the map is expressed through two harmonic conjugate functions evaluated spectrally.

    \item [ii.] A quantitative assessment of the failure of linear theory at high Froude number. The nonlinear leading crest is substantially higher than its linear counterpart, the onshore and offshore wave trains become strongly asymmetric, and the solution loses the symmetry $A \mapsto -A$ that linear theory enforces.

    \item [iii.] Computation of the free-surface profile and of the underlying velocity potential for arbitrary laboratory topographies through Hermite interpolation, with validation against two independent sets of experiments and a quantitative error assessment.

    \item [iv.] An investigation of the influence of the obstacle shape on the generated waves within the full Euler equations, at fixed area, amplitude and centre of mass, showing that the shape is not a primary factor.

    \item [v.] An  numerical study  of particle dynamics at the free surface and inside the fluid, including the dependence of the horizontal and vertical displacements on the initial depth and on the initial horizontal position.
\end{enumerate}

The remainder of this paper is organized as follows. Section~\ref{Formulation} presents the mathematical formulation, including the governing equations, the conformal mapping, the trajectory system for non-autonomous flows and
the interpolation procedure. Section~\ref{benchmark} validates the methodology against two sets of laboratory experiments. Section~\ref{results} contains the numerical results: the effects of the Froude number, the acceleration and the amplitude on the generated wave, the influence of the obstacle shape, the structure of the velocity field and the particle dynamics in three regions of the fluid domain.
Section~\ref{Conclusion} summarizes our findings.


\section{Formulation}
\label{Formulation}

\subsection{Dimensionless Euler equations}

Consider an ideal fluid with incompressible and irrotational flow in a finite-depth channel. The seabed is time-dependent and given by $h_0 + h(x,t)$, whereas the free surface is denoted by ${\zeta}(x, t)$. We assume that $h_0$ is a characteristic depth and use it as the length scale, both vertically and horizontally. The velocity and time scales are taken as $(gh_0)^{1/2}$ and $(h_0/g)^{1/2}$, respectively, where $g$ is the acceleration of gravity. Therefore, we obtain the Euler equations in their dimensionless form
        \begin{align}
         \label{eq:eu1}
            & \Delta{{\phi}}= 0 \;\  \mbox{for} \;\ -1+h(x,t) < y < \zeta (x,t), \\
            \label{eq:eu2}
            & {{\phi}}_{y} =h_{t}+{\phi}_{x}h_{x} \;\ \mbox{at} \;\ y = -1+h(x,t), \\
            \label{eq:eu3}
            & {\zeta}_{t} + \phi_x{{\zeta}}_{x}-{{\phi}}_{y}=0
            \;\ \mbox{at} \;\ y = \zeta (x,t), \\
            \label{eq:eu4}
            & {{\phi}}_{t}+  \frac{1}{2}(\phi_x^2 + \phi_y^2)+{{\zeta}}= 0\;\ \mbox{at} \;\ y = \zeta (x,t).
               \end{align}
These equations are complemented by periodic boundary conditions with period $2L$. 

System \eqref{eq:eu1}--\eqref{eq:eu4} is solved numerically by the conformal mapping method described in the next section. Conformal maps have a long history in the numerical study of nonlinear water waves, starting from Dyachenko et al. \cite{Dyachenko:1996} for a flat seabed and extended by Ruban \cite{Ruban:2005} to time-dependent bottoms; further applications and recent developments can be found in \cite{Wilkening:2023} and the references therein.


\subsection{Conformal mapping}

A double conformal mapping is employed to flatten both the free surface wave and the seabed. This approach transforms the physical domain into a canonical domain, effectively serving as a bridge between them. The canonical domain is represented by a rectangle, allowing the Euler equations to be rewritten as a system of ordinary differential equations. More specifically, we consider $\Omega$ the instantaneous fluid domain, i.e. the fluid domain at a fixed value of $t$,
\begin{align*}
    \Omega = \{x+iy\in\mathbb{C}\mid -1 + h(x,t) \leq y\leq \zeta(x,t)\ \text{and}\ -L\leq x\leq L\}
\end{align*}
and obtain a conformal map from a uniform strip $\widetilde{\Omega}$ of length $2L$ and width $D$ onto $\Omega$:
\begin{align*}
    f:\{\xi + i\eta\in\mathbb{C}\mid -D(t)\leq\eta \leq 0\ \text{and}\ -L\leq\xi\leq L \}\longrightarrow \Omega,
\end{align*}
whose components are given by 
\begin{align*}
    f(\xi + i\eta,t) = X(\xi,\eta,t) + iY(\xi,\eta,t).
\end{align*}

We denote the Fourier coefficients by 
\begin{align*}
    \mathcal{F}_{k_j}[g(\xi)]=\hat{g}(k_j)=\frac{1}{2L}\int_{-L}^{L}g(\xi)e^{-ik_j\xi}\,d\xi,
\end{align*}
and the inverse Fourier transform by 
\begin{align*}
    \mathcal{F}^{-1}_{k_j}[\hat{g}(k_j)](\xi)=g(\xi)=\sum_{j=-\infty}^{\infty}\hat{g}(k_j)e^{ik_j\xi},
\end{align*}
with $k_j=(\pi/L)j$, $j\in\mathbb{Z}$. 

Let
\begin{align*}
   X(\xi,0,t) = \mathbf{X}(\xi,t)\quad \text{and}\quad Y(\xi,0,t) = \mathbf{Y}(\xi,t).
\end{align*}
We impose that they parametrize the free surface $(x,\zeta(x,t))$, so $\mathbf{Y}(\xi,t) = \zeta(\mathbf{X}(\xi,t),t)$ is satisfied for all $\xi$ and all $t$. For the bottom coordinates, we use the following notation
\begin{align*}
   X(\xi,-D,t) = \mathbf{X}_b(\xi,t)\quad \text{and}\quad Y(\xi,-D,t) = -1 + \mathbf{H}(\xi,t),
\end{align*}
where $\mathbf{H}(\xi,t) = h(\mathbf{X}_b(\xi,t),t)$.
Furthermore, let $\mathbf{\Phi}(\xi,t)$ and $\mathbf{\Psi}(\xi,t)$ represent the surface velocity potential 
and its harmonic conjugate in the new variables. The components of the map can then be written as
\begin{align}
    X(\xi,\eta,t) = & \mathcal{F}^{-1}_{k \neq 0}\left[\dfrac{i\coth{(kD)}\cosh{(k\eta)}\widehat{\mathbf{H}}(k,t)}{\cosh{(kD)}}\right] + \mathcal{F}^{-1}_{k \neq 0}\left[\dfrac{-i\cosh{(k(D + \eta))}\widehat{\mathbf{Y}}(k,t)}{\sinh{(kD)}}\right] + \xi, \label{system_xy_clean_x} \\
         Y(\xi,\eta,t) = & \mathcal{F}^{-1}_{k \neq 0}\left[-\dfrac{\coth{(kD)}\sinh{(k\eta)}\widehat{\mathbf{H}}(k,t)}{\cosh{(kD)}}\right] + \mathcal{F}^{-1}_{k \neq 0}\left[\dfrac{\sinh{(k(D + \eta))}\widehat{\mathbf{Y}}(k,t)}{\sinh{(kD)}}\right] + \widehat{\mathbf{Y}}(0,t) + \eta, \label{system_xy_clean_y}
\end{align}
 while the velocity potential in canonical coordinates, $\bar{\phi}(\xi,\eta,t) = \phi(X(\xi,\eta,t),Y(\xi,\eta,t),t)$, is given by
\begin{equation}\label{eq:Pot1}
    \bar{\phi}(\xi,\eta,t) = \mathcal{F}^{-1}_{k\ne 0}\bigg[\frac{\cosh(k(\eta+D))\widehat{\mathbf{\Phi}}}{\cosh(kD)}
    +\frac{\sinh(k\eta)}{k\cosh(kD)}\widehat{h_{t}\mathbf{{X_{b}}_{\xi}}}\bigg] + \widehat{\mathbf{\Phi}}(0,t)  + {\widehat{h_{t}\mathbf{{X_{b}}_{\xi}}}(0,t)}\eta.
\end{equation}
 Remark that, through Cauchy Riemann equations, we can write $\bar{\psi}(\xi,\eta,t) = \psi(X(\xi,\eta,t),Y(\xi,\eta,t),t)$, the harmonic conjugate of $\bar{\phi}$, through $\bar{\psi}_{\xi}(\xi,\eta,t) = -\bar{\phi}_{\eta}(\xi,\eta,t)$. 

From these definitions, the following system governing the evolution of the free surface wave can be derived,  which constitutes the core of the proposed method
\begin{align}
    \mathbf{Y}_t &= \mathbf{Y}_{\xi} \mathcal{C}\left[\dfrac{\mathbf{\Psi}_{\xi}(\xi,t)}{J} \right] - \mathbf{X}_{\xi}{\left(\dfrac{\mathbf{\Psi}_{\xi}(\xi,t)}{J}\right)}.\label{eq_Y_t} \\
    \label{systemYPhi}
    \mathbf{\Phi}_t &= \mathcal{C}\left[\dfrac{\mathbf{\Psi}_{\xi}(\xi,t)}{J} \right]\mathbf{\mathbf{\Phi}}_{\xi} - \frac{1}{2J}(\mathbf{\mathbf{\Phi}}_{\xi}^2 - \mathbf{\mathbf{\Psi}}_{\xi}^2) - \mathbf{Y},
\end{align}
where $\mathcal{C}\left[ \frac{\mathbf{\Psi}_{\xi}}{J}\right]$ is given by
\begin{equation}\label{operator_C}
     \mathcal{C}\left[ \frac{\mathbf{\Psi}_{\xi}}{{J}}\right] = 
             \mathcal{F}^{-1}_{k_j \neq 0}\left[i \coth(k_jD)\left(\frac{1}{\cosh{(kD)}}\mathcal{F}\left[\frac{h_t\mathbf{X}_{b_{\xi}}}{J_b} \right] + \mathcal{F}\left[\frac{\mathbf{\Psi}_{\xi}(\xi,t)}{J}\right]\right)\right]  - \widehat{M}(0,t),
\end{equation}
$J = \mathbf{X}_{{\xi}}^2 + \mathbf{Y}_{{\xi}}^2$ and $J_b = \mathbf{X}_{b_{\xi}}^2 + \mathbf{Y}_{b_{\xi}}^2$ are the Jacobian evaluated at $\eta = 0$ and $\eta = -D$, respectively, $h_t = h_t(\mathbf{X}_{b_{\xi}},t)$ and
\begin{equation*}
    M(\xi,t) = \mathbf{X}_\xi \mathcal{F}^{-1}_{k_j \neq 0}\left[i \coth(k_jD) \left(\frac{1}{\cosh{(kD)}}\mathcal{F}\left[\frac{h_t\mathbf{X}_{b_{\xi}}}{J_b} \right] + \mathcal{F}\left[\frac{\mathbf{\Psi}_{\xi}(\xi,t)}{J}\right]\right)\right] +  \mathbf{Y}_\xi \frac{\mathbf{\Psi}_{\xi}}{J}.
\end{equation*}
We set $D(t) = 1 + \widehat{\mathbf{Y}}(0,t) - \widehat{\mathbf{H}}(0,t)$ to guarantee that both canonical and physical domain have the same horizontal length.

The functions $\mathbf{X}(\xi,t)$ and $\mathbf{\Psi}(\xi,t)$ are computed through
\begin{align*}
         \mathbf{X}(\xi,t) & = \xi+ \mathcal{F}^{-1}_{k_j\ne 0}\bigg[\frac{i\coth(k_jD)\widehat{\mathbf{H}}}{\cosh(k_jD)}\bigg]  +\mathcal{F}^{-1}_{k_j\ne 0}\bigg[-i\coth(k_jD)\widehat{\mathbf{Y}}\bigg] \\
        \mathbf{\Psi}_{\xi}(\xi,t) & =\mathcal{F}^{-1}\Bigg[i\tanh(k_{j}D) \widehat{\mathbf{\Phi}_{\xi}}-\frac{\widehat{h_{t}x_{b\xi}}}{\cosh(k_{j}D)} \Bigg]
\end{align*}
and  $\mathbf{X}_{b}(\xi,t)$ is found implicitly, since $\mathbf{H}(\xi,t) = h(\mathbf{X}_b,t)$, i.e., 
\begin{align}
\label{xbxi}
             \begin{aligned}
        \mathbf{X}_{b}(\xi,t)& =\xi + \mathcal{F}^{-1}_{k_j\ne 0}\bigg[i\tanh(k_jD)\widehat{\mathbf{H}}\bigg] + \mathcal{F}^{-1}_{k_j\ne 0}\bigg[i\coth(k_jD)\bigg[\frac{\widehat{\mathbf{H}}}{\cosh^2(k_jD)} - \frac{\widehat{\mathbf{Y}}}{\cosh{(k_jD)}} \bigg]\bigg].  
        \end{aligned}
          \end{align}

Derivatives along the horizontal direction are computed spectrally via the {Fast Fourier Transform (FFT)}, and the system \eqref{eq_Y_t}-\eqref{systemYPhi} is solved using the classical { fourth-order Runge-Kutta method (RK4)}.
Once the solution $\mathbf{Y}(\xi,t)$ is obtained in the canonical domain, the free surface profile in the physical space, is given by $(\mathbf{X}(\xi,t),\mathbf{Y}(\xi,t)).$ Further details on this technique can be found in the work of Poletto et al. \cite{Poletto:2025}.

With the formulas for the coordinates of the conformal map and the velocity potential at hand, it is possible to determine the velocity field in the original fluid domain as a function of $\xi,\eta$. Indeed, by means of the chain rule, we can express them by
 \begin{align}
    {\phi}_x & = \dfrac{1}{X_{\xi}^2 + Y_{\xi}^2}(\bar{\phi}_{\xi}X_{\xi} + \bar{\psi}_{\xi}Y_{\xi}) \label{phi_x}, \\
    \phi_y & = \dfrac{1}{X_{\xi}^2 + Y_{\xi}^2}(\bar{\phi}_{\xi}Y_{\xi} - \bar{\psi}_{\xi}X_{\xi})\label{phi_y}. 
\end{align}

Once the free surface wave dynamics is computed through formulas \eqref{eq_Y_t} and \eqref{systemYPhi}, the velocity field can be computed using equations \eqref{system_xy_clean_x}-\eqref{phi_y}.



\subsection{Particle dynamics in the fluid}

Let $P(t) = (x(t),y(t))$ denote the trajectory of a fluid particle that starts its motion at $(x_0,y_0)$. The trajectory is governed by the following system of ODE:
\begin{equation*}\label{sistema_trajetoria}
    \begin{cases}
    \dfrac{dx}{dt}(t) = \phi_x(x(t),y(t),t), \\
    \dfrac{dy}{dt}(t) = \phi_y(x(t),y(t),t), \\
    (x(0),y(0)) = (x_0,y_0).
    \end{cases}
\end{equation*}
Our approach however, gives no direct knowledge of the function $\phi$ in the physical domain. Instead, it gives us the values of $\tilde{\phi}$, the velocity potential in the canonical domain, in terms of the canonical variables $\xi$ and $\eta$. 
So, given a point $\tilde{P}_0 = (\xi_0,\eta_0)$ in the canonical domain, we want to determine its trajectory $\tilde{P}(t) = ({\xi}(t),{\eta}(t))$ in the canonical domain and then, recover the corresponding trajectory in the physical fluid doamin.

Note that we can write
\begin{equation*}
    (x(t),y(t)) = (X(\xi(t),\eta(t),t),Y(\xi(t),\eta(t),t)),
\end{equation*}
so
\begin{equation*}
\begin{cases}
    \dfrac{dx}{dt} = X_{\xi}\dfrac{d \xi}{dt} + X_{\eta}\dfrac{d \eta}{dt} + X_t,\vspace{1mm} \\ 
    \dfrac{dy}{dt} = Y_{\xi}\dfrac{d \xi}{dt} + Y_{\eta}\dfrac{d \eta}{dt} + Y_t. 
\end{cases}
\end{equation*}
Using $dx/dt = \phi_x$ and $dy/dt = \phi_y$, we invert this system and obtain
\begin{align*}
    \dfrac{d \xi}{d t} & = \frac{1}{J}\left(Y_{\eta}\phi_x - X_{\eta}\phi_y - Y_{\eta}X_t + X_{\eta}Y_t \right), \\
    \dfrac{d \eta}{dt} & = \frac{1}{J}\left(-Y_{\xi}\phi_x + X_{\xi}\phi_y + Y_{\xi}X_t - X_{\xi}Y_t\right).
\end{align*}
Considering equations \eqref{phi_x} and \eqref{phi_y} it follows that
\begin{align*}
    \dfrac{d\xi}{dt} & = \frac{\tilde{\phi}_{\xi}}{J} - \frac{1}{J} \left(X_{\xi}X_t + Y_{\xi}Y_t \right).
\end{align*}
In a similar manner, we can write
\begin{equation*}
    \dfrac{d \eta}{dt} = \frac{\tilde{\phi}_{\eta}}{J} - \frac{1}{J}\left(X_{\xi}Y_t - Y_{\xi}X_t\right).
\end{equation*}

Therefore, we reached the following system for the particle trajectories in the canonical domain
\begin{equation}\label{system_particle_canonical}
    \begin{cases}
        \dfrac{d \xi}{d t} = \dfrac{\tilde{\phi}_{\xi}(\xi,\eta,t)}{J(\xi,\eta,t)} - \Gamma(\xi,\eta,t), \\
        \dfrac{d\eta}{d t} = \dfrac{\tilde{\phi}_{\eta}(\xi,\eta,t)}{J(\xi,\eta,t)} - \Lambda(\xi,\eta,t), \\
        \xi(0),\eta(0) = (\xi_0,\eta_0)
    \end{cases}
\end{equation}
where $\Gamma$ and $\Lambda$ are the real and imaginary part of the following analytic function
\begin{equation*}\label{analytical_function}
    w(\xi,\eta,t) = \frac{X_t + iY_t}{X_{\xi} + iY_{\xi}} = \frac{X_tX_{\xi} + Y_tY_{\xi}}{X_{\xi}^2 + Y_{\xi}^2} + i\frac{Y_tX_{\xi} - X_tY_{\xi}}{X_{\xi}^2 + Y_{\xi}^2}.
\end{equation*}
The formulas for both functions are
\begin{equation}\label{Gamma_equation}
    \Gamma(\xi,\eta,t) = \mathcal{F}^{-1}_{k\neq 0} \left[-i\frac{\cosh{(k(D + \eta))}}{\sinh{(kD)}}\widehat{f}(k,t) +i \frac{\cosh{(k\eta)}}{\sinh{(kD)}}\widehat{g}(k,t)\right] + \left( \frac{\widehat{f}(0,t)}{D} - \frac{\widehat{g}(0,t)}{D}\right)\xi,
\end{equation}
and
\begin{equation}\label{Lambda_equation}
    \Lambda(\xi,\eta,t) = \mathcal{F}^{-1}_{k \neq 0}\left[\frac{\sinh{(k(D + \eta))}}{\sinh{(kD)}}\widehat{f}(k,t) - \frac{\sinh{(k\eta)}}{\sinh{(kD)}}\widehat{g}(k,t) \right] + \frac{(D + \eta)}{D}\widehat{f}(0,t) - \frac{\eta}{D}\widehat{g}(0,t)
\end{equation}
where
\begin{equation*}
     f(\xi,t) = -\frac{\mathbf{\Psi}_{\xi}(\xi,t)}{J} \quad \text{ and } \quad g(\xi,t) = \frac{(h_tX_{b\xi})(\xi,t)}{J_b}.
\end{equation*}
The derivation of both formulas is given in Appendix \ref{Gamma_Lambda_derivation}.

Two features of system \eqref{system_particle_canonical} deserve attention. First, the system is posed entirely in canonical coordinates: no inversion of the conformal map and no interpolation of the velocity field onto the particle position are required, and every term is evaluated spectrally on the same grid used to advance the free surface in time. Second, the right-hand side depends explicitly on $t$ through $\Gamma$ and $\Lambda$, which encode the motion of the map itself. As mentioned earlier, previous computations of particle paths under Dyachenko-type maps \cite{Chen:2021,Ige:2024,Nachbin_RibeiroJr:2014,Nachbin_RibeiroJr:2017} assume a periodic traveling wave of permanent form. In a frame moving with the wave the map becomes stationary, $X_t = Y_t = 0$, so that $\Gamma \equiv \Lambda \equiv 0$ and \eqref{system_particle_canonical} reduces to an autonomous planar system whose orbits can be analysed with phase-plane techniques. No such frame exists for the flows considered here: the obstacle accelerates from rest, translates and decelerates according to a prescribed law, the wave field is transient, and the conformal map is genuinely time dependent. Formulas \eqref{Gamma_equation} and \eqref{Lambda_equation} remove this obstruction, so that system \eqref{system_particle_canonical} applies to arbitrary unsteady potential flows over a moving seabed.

\subsection{Handling laboratory topographies}



Handling realistic experimental setups requires some technical care as it involves computing the conformal map on the actual fluid domain. This needs solving the implicit relation $\mathbf{H}(\xi,t) = h(\mathbf{X}_b,t)$ with $\mathbf{X}_b$ given by equation \eqref{xbxi}. As shown in Poletto et al. \cite{Poletto:2025}, a numerical fixed point iteration suffices, and its convergence to a desired tolerance is typically fast. However, for the iteration scheme to work, it is necessary to have an exact formula for moving topography, i.e. the function $h(x,t)$ must be known explicitly. In  laboratory experiments, moving obstacles are often given by rectangular, triangular, or elliptical shapes so, we introduce the following technique to get the corresponding function $h(x,t)$. 

Throughout this paper we only consider seabed motions that are rigid translations of its initial profile, so $h(x,t) = h(x - s(t))$, where $s(t)$ is the position of the centre of mass of the obstacle. We emphasize that this velocity is not required to be constant in time, in contrast to the assumption made, for instance, in \cite{Wang:2019}. The initial configuration $h(x)$ is then obtained through a { piecewise cubic Hermite interpolating polynomial (PCHIP)}. We specify the number of points from the original topography used for the interpolation and construct the corresponding polynomial. This procedure ensures an accurate representation of the initial seabed shape, since the interpolation reproduces both the function $h(x)$ and its derivative. 
A somewhat similar approach was used by Zhao et al. \cite{Zhao:2026}, through a convolution-based method, that smooths out the obstacle's shape and produces a function $h(x,t)$ that is then used with the method described herein. Both methods share the advantage of reproducing controlled physical conditions as those used in laboratory experiments. The next section demonstrates the effectiveness of this approach.


\section{Benchmark} \label{benchmark}

In this section, we validate our numerical method and the physical model by direct comparison between numerical results with experimental measurements and demonstrate that the model reproduces realistic scenarios accurately. From here on, FFT computations employ $N = 2^{10}$ modes for the benchmarking tests and $N = 2^{11}$ modes for the numerical simulations. Time integration is performed using the Runge--Kutta method with a fixed time step of $0.01$. The fluid domain is chosen with dimensionless length of approximately 84,  equivalent to 84 times the depth of the channel. 

Aliasing errors are controlled by a low-pass filter applied at every time step, in which the Fourier modes with $k \geq N/3$ are discarded. The same filter is used in all the simulations reported in this work.




\subsection{Model validation against experimental data}

\begin{figure}[!htb]
    \centering
    \includegraphics[scale=1]{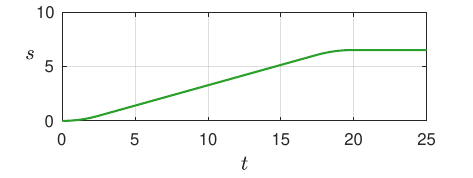}
    \includegraphics[scale=1]{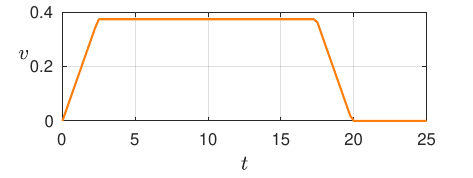}
    \caption{Center mass position and velocity for the semi-elliptical topography moving horizontally, according to parameters set in Table \ref{Dimensional_vs_dimensionless_bechmark} for $Fr = 0.375$.}
    \label{grafico_s_e_v}
\end{figure}

In order to validate our procedure, we take runs for Whittaker's experiment \cite{Whittaker:2015}, where the author considers a semi-elliptical block moving horizontally. This block is described by the function
\begin{equation*}
    h(x,t) = \begin{cases}A \sqrt{1 - 4{\left(\frac{(x - s(t))}{L_{b}}\right)}^2}, \quad \text{ if } \quad |x - s(t)| \leq \frac{L_b}{2} \\
    0, \quad \text{ if } \quad |x - s(t)| > \frac{L_{b}}{2}
    \end{cases},
\end{equation*}
where $A$ and $L_b$ are the height and width of the block, respectively given by $A = 0.1486$ and $L_b = 2.8571$, and $s(t)$ is given by
\begin{equation}\label{s_function}
    s(t) = \begin{cases}
        \frac{a_0 \cdot t^2}{2}, \quad \text{ if } \quad 0 \leq t < t_{1}, \\
        \frac{a_0 \cdot t_1^2}{2} + u_t\cdot (t - t_{1}), \quad \text{ if } \quad t_{1} \leq t < t_{2}, \\
        \frac{a_0 \cdot t_1^2}{2} + u_t\cdot (t_2 - t_{1}) -\frac{a_0{(t - t_{2})}^2}{2}, \quad \text{ if } \quad t_{2} \leq t < t_{3}, \\
        \frac{a_0 \cdot t_1^2}{2} + u_t\cdot (t_{2} - t_{1}) -\frac{a_0{(t_{3} - t_{2})}^2}{2}, \quad \text{ if } \quad t \geq t_{3}.
    \end{cases}
\end{equation}
For benchmarking purposes, we adopt runs 6, 12, and 18 from Whittaker et al.~\cite{Whittaker:2015}, 
whose parameters for times $t_1,t_2$ and $t_3$ are given by Table \ref{Dimensional_vs_dimensionless_bechmark}.  They are chosen so that the block accelerates uniformly from rest, travels at constant velocity, and then decelerates at the same magnitude of acceleration until it stops. The parameter $u_t$ is the terminal velocity. In our scaling, it corresponds to the {Froude number} at it will be referred to as $Fr$. The position and velocity of the center of mass are shown in Figure \ref{grafico_s_e_v}.
\begin{table}[!htb]
    \centering
    \begin{tabular}{c|c|c}\hline
        Parameter & Dimensional value & Dimensionless value 
        \\ \hline 
        $F_r$ & $0.164 \, m/s$ & $0.125$  \\
        $t_1$ & $0.108 \, s$ & $0.811$  \\ 
        $t_2$ & $2.110 \, s$ & $15.787$  \\ 
        $t_3$ & $2.220 \, s$ & $16.616$  \\ 
        \hline 
        $F_r$ & $0.328 \, m/s$ & $0.250$  \\
        $t_1$ & $0.219 \, s$ & $1.640$  \\ 
        $t_2$ & $2.220 \, s$ & $16.616$  \\ 
        $t_3$ & $2.437 \, s$ & $18.238$ \\ \hline
        $F_r$ & $0.492 \, m/s$ & $0.375$  \\
        $t_1$ & $0.327 \, s$ & $2.451$  \\ 
        $t_2$ & $2.329 \, s$ & $17.427$  \\ 
        $t_3$ & $2.656 \, s$ & $19.878$ \\ \hline
    \end{tabular}
    \caption{Comparison between dimensional parameters of the experiments(runs 6, 12 18) by Whittaker et al. \cite{Whittaker:2015} and our dimensionless scale. All regimes occur with $a_0 = 1.5 \, m/s^2$ which corresponds to $a_0 = 0.153$.}
    \label{Dimensional_vs_dimensionless_bechmark}
\end{table}

\begin{figure}[!htb]
    \centering
    \includegraphics[scale=1]{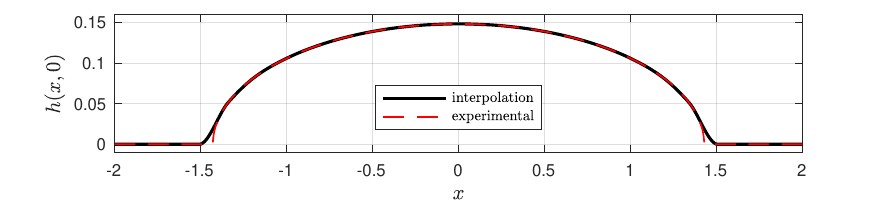}
     \caption{Interpolated and experimental topography, for experiment of Whittaker et al. \cite{Whittaker:2015}}
    \label{topographies_benchmark}
\end{figure}

In order to obtain a good quality of the interpolation, we fix the number of interpolation points at 501. We will analyze the three cases presented by Whittaker et al., corresponding to Froude numbers 0.125, 0.25, and 0.375, and examine the behavior of the generated waves. The comparison of the topographies in these three cases is shown in Figure \ref{topographies_benchmark}. 

\begin{figure}[!ht]
    \centering
    \includegraphics[scale = 1]{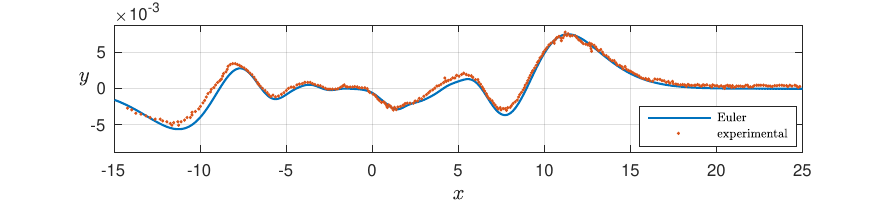}
    \includegraphics[scale = 1]{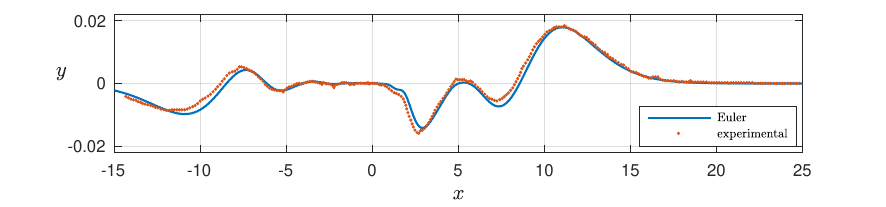}
    \includegraphics[scale = 1]{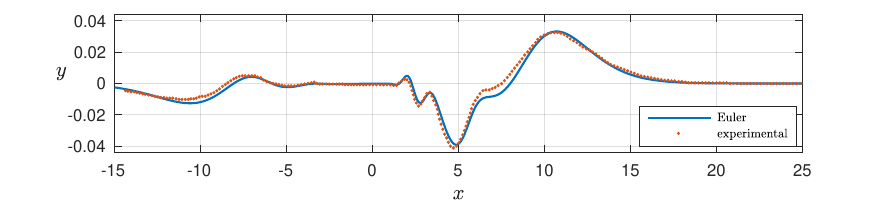}
    \caption{Waves generated for the Hermite interpolated topography for 501 points, compared to the experimental data, for three different Froude numbers (top: 0.125, middle: 0.25 and bottom: 0.375), considering $t = 13.5$.}
    \label{benchmark_waves}
\end{figure}

We can see that the interpolation provides sufficient accuracy when compared with the laboratory geometry. The generated waves are presented in Figure \ref{benchmark_waves}. Overall, our model predicts the position of both crest and trough with remarkable accuracy, and the wave amplitude and dispersion are also well represented. Our results agree better with experimental data compared to simpler models \cite{Firdaus:2025,Firdaus:2025_2,Whittaker:2015}, while small differences may be attributed to several factors, including the approximation of gravity acceleration and other experimental parameters; wave reflections within the tank; filling taps located at the end of the flume; departures of the experimental configuration from the idealized numerical geometry; sidewall friction; and finally, the fluid exchange through the slot in the false floor \cite{Whittaker:2015}.
 
\begin{figure}[!ht]
    \centering
    \includegraphics[scale = 1]{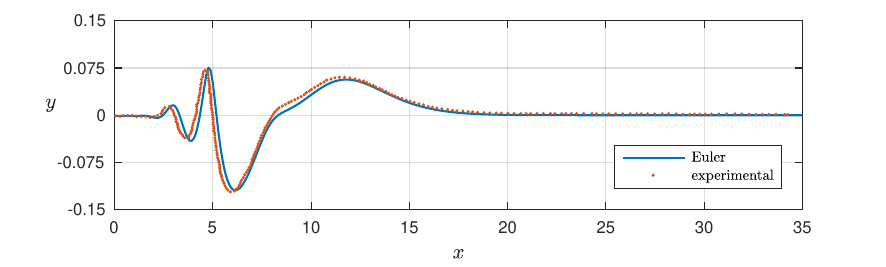}
    \includegraphics[scale = 1]{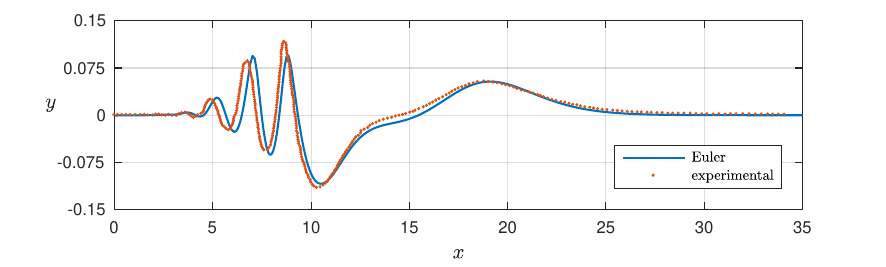}
   \caption{Waves generated for the Hermite interpolated topography for 501 points, compared to the experimental data for $Fr = 0.5$, considering $t = 14.98$ (top) and $t = 22.46$ (bottom).}
    \label{benchmark_waves_Whit_17_case3}
\end{figure}

Next, we compare our model with a second set of experiments reported by Whittaker et al. \cite{Whittaker:2017}. Using the same experimental apparatus, the authors investigated the behavior of waves induced by horizontal landslides at higher Froude numbers. Specifically, the analysis was conducted for values from $Fr = 0.5$ up to $Fr \leq 0.75$. The corresponding parameters, including acceleration, velocity, and the time intervals associated with changes in the landslide motion for $Fr = 0.5$, are listed in Table \ref{Dimensional_vs_dimensionless_2}. The results for this case are presented in Figure \ref{benchmark_waves_Whit_17_case3}, which shows again a good agreement with experimental data.
\begin{table}[!htb]
    \centering
    \begin{tabular}{c|c|c}\hline
        Parameter & Dimensional value & Dimensionless value \\ \hline
        $a_0$ & $1.5 \, m/s^2$ & $0.153$  \\
        $u_t$ & $0.655 \, m/s$ & $0.5$  \\
        $t_1$ & $0.438 \, s$ & $3.279$  \\ 
        $t_2$ & $2.437 \, s$ & $18.238$  \\ 
        $t_3$ & $2.875 \, s$ & $21.518$  \\ \hline
    \end{tabular}
    \caption{Comparison between dimensional parameters of the experiments (run 6) by Whittaker et al. \cite{Whittaker:2017} and our dimensionless scale for the experiment relative to Froude number $0.5$.}
    \label{Dimensional_vs_dimensionless_2}
\end{table}


We also quantify the deviation of our results from the laboratory measurements through
\begin{equation}\label{error_Whittaker}
    \mathcal{E} = \Vert \zeta_w - \zeta\Vert_2,
\end{equation}
where $\zeta_w$ and $\zeta$ denote the Whittaker's experimental data and our nonlinear solution, respectively. Spline interpolation is applied to the numerical solution to ensure that both quantities are evaluated at the same spatial points. The results are displayed in Table \ref{Norm_deviation}.
\begin{table}[!htb]
    \centering
    \begin{tabular}{c|c}\hline
        Case & $\mathcal{E}$ \\ \hline
        $1$ & $7.2480 \cdot 10^{-6}$ \\
        $2$ & $3.7532 \cdot 10^{-5}$   \\
        $3$ & $8.6028 \cdot 10^{-5}$   \\ 
        $4$ & $1.0413 \cdot 10^{-2}$  \\ 
        $5$ & $2.3101 \cdot 10^{-2}$ \\ \hline
    \end{tabular}
    \caption{Deviation of the nonlinear solution from Whittaker's experimental dataset for different test cases, computed by formula \eqref{error_Whittaker}. Cases 1--3 correspond to the configurations shown in Figure \ref{benchmark_waves}, with $Fr=0.125$, $Fr=0.25$, and $Fr=0.375$, respectively. Cases 4 and 5 correspond to the configurations shown in Figure \ref{benchmark_waves_Whit_17_case3}, for $Fr = 0.5$ at $t=14.98$ and $t=22.46$, respectively.}
    \label{Norm_deviation}
\end{table}

With the model thus validated, we proceed to investigate the departures from linearity in different regimes characterised by different values of the Froude number.


\section{Numerical results} \label{results}

A major advantage of the present numerical approach is its ability to explore a wider range of parameters than is feasible in laboratory experiments. For this reason, throughout this section we consider a setup similar to that proposed by Whittaker et al. \cite{Whittaker:2015}, as described in the previous section. Specifically, we retain the same topographic profile (with $A = 0.1486$) and compare two cases: $Fr = 0.125$, corresponding to the experiment of Whittaker et al., and $Fr = 1.2$. 
The corresponding that determine the block motion are given in Table \ref{Dimensional_vs_dimensionless_exploration}. Considering Froude numbers greater than 1.2 did not lead to significant qualitative changes in the generated wave field. Therefore, we focus our analysis on these Froude numbers.
\begin{table}[!htb]
    \centering
    \begin{tabular}{c|c|c}\hline
        Parameter & Dimensional value & Dimensionless value 
        \\ \hline 
        $F_r$ & $0.164 \, m/s$ & $0.125$  \\
        $t_1$ & $0.108 \, s$ & $0.811$  \\ 
        $t_2$ & $2.110 \, s$ & $15.787$  \\ 
        $t_3$ & $2.220 \, s$ & $16.616$  \\ 
        \hline 
        $F_r$ & $1.574 \, m/s$ & $1.2$  \\
        $t_1$ & $1.048 \, s$ & $7.843$  \\ 
        $t_2$ & $3.049 \, s$ & $22.819$  \\ 
        $t_3$ & $4.0973 \, s$ & $30.662$ \\ \hline
    \end{tabular}
    \caption{Comparison between dimensional parameters and our dimensionless scale in the exploration of nonlinear effects. Both cases occur with $a_0 = 1.5 \, m/s^2$ with correponds to $a_0 = 0.153$.}
    \label{Dimensional_vs_dimensionless_exploration}
\end{table}

\begin{figure}[!htb]
         \centering
         \includegraphics[scale=0.4]{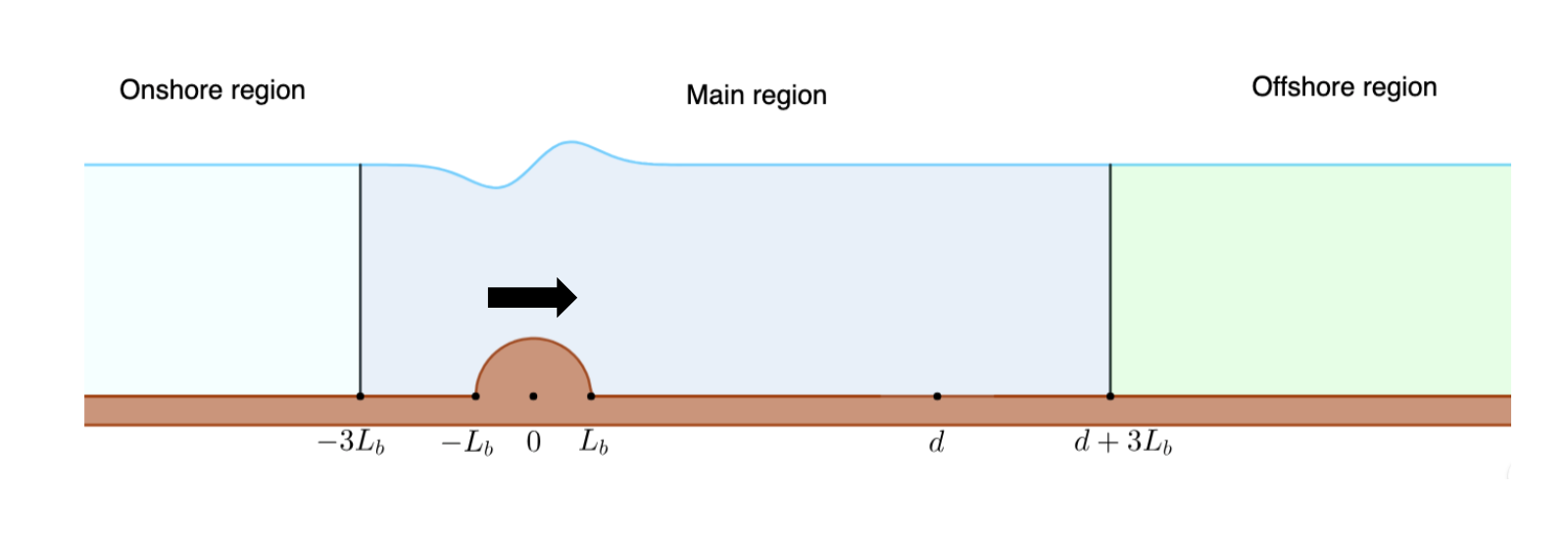}
        \caption{Schematic representation of the three domain fluid regions The block length is $2L_b$ and the distance traveled is $d$.}
        \label{schematic_figure_particles}
\end{figure}

Initially, inspired by the nomenclature of Whittaker et al. \cite{Whittaker:2015}, we divide the fluid domain in three regions, called herein as \textit{onshore, main} and \textit{offshore region}. See Figure \ref{schematic_figure_particles}. The main region corresponds to the portion of the fluid domain located above the block trajectory, extended by one block length on both sides of the slide path.

For the following results, departure from linearity is studied by comparing the solution of the nonlinear model against a linear, fully dispersive surface wave
\begin{equation*} 
    \zeta(x,t) =  \mathcal{F}^{-1}\bigg[\text{sech}(k_j)\int_0^t \cos(\omega_j(t-s))\hat{h}_t(k_j,t) ds\bigg],
    \label{eta1}
\end{equation*}
where $\omega_j^2 = k_j\tanh{(k_j)}$. Details about the derivation of the formula can be found in the work of Jing et al \cite{Jing:2020} or Lo et al \cite{Lo:2017}.

According to Whittaker et al.~\cite{Whittaker:2017}, wave breaking occurs for $Fr = 0.625$ and above. However, for Froude numbers equal to or slightly greater than unity, breaking does not occur when other parameters are kept the same, which allows us to explore this regime in the present work. Such high Froude numbers are less common but not unprecedented in this context; see, for instance, Renzi et al.~\cite{Renzi:2023} and Wang et al.~\cite{Wang:2019}, where values of the Froude number equal to or greater than unity were also considered to investigate nonlinear effects. 


\subsection{Exploring nonlinear effects on the generated wave}

To account for differences between the linear model and the nonlinear one, we focus the shape of the wave profile and its symmetry, at different values of the Froude number. Then, we examine the effects of acceleration and topographic amplitude on the resulting wave patterns. To obtain clearer results, we doubled the computational domain while preserving the spatial resolution by also doubling the number of discretization points ($2^{11}$).
\begin{figure}[!htb]
         \centering
         \includegraphics[scale=1]{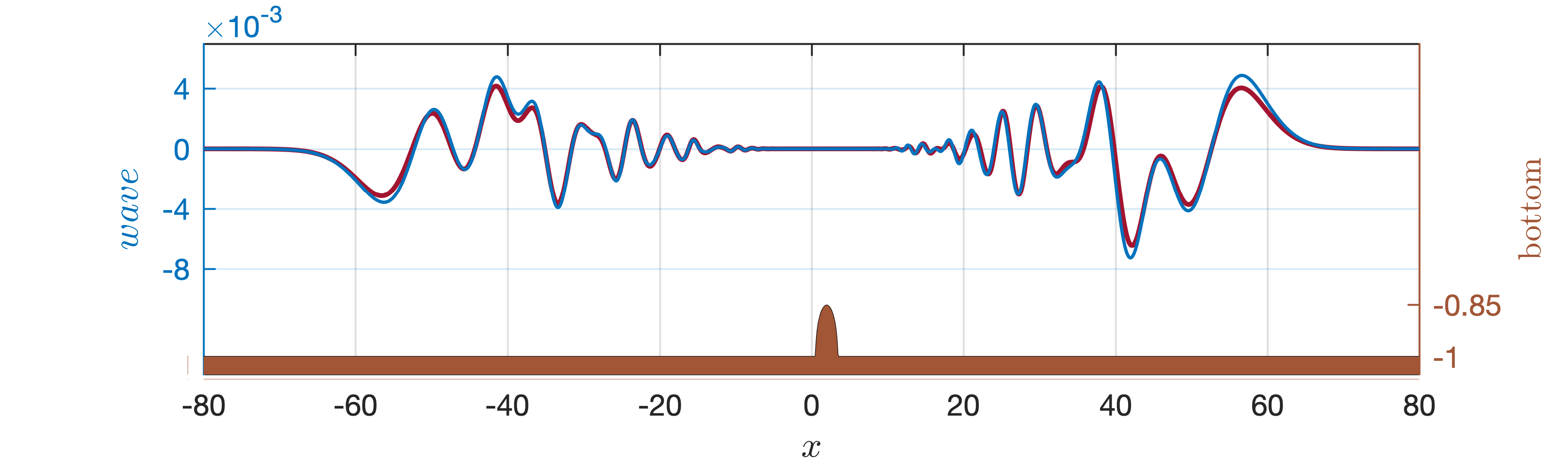}
         \includegraphics[scale=1]{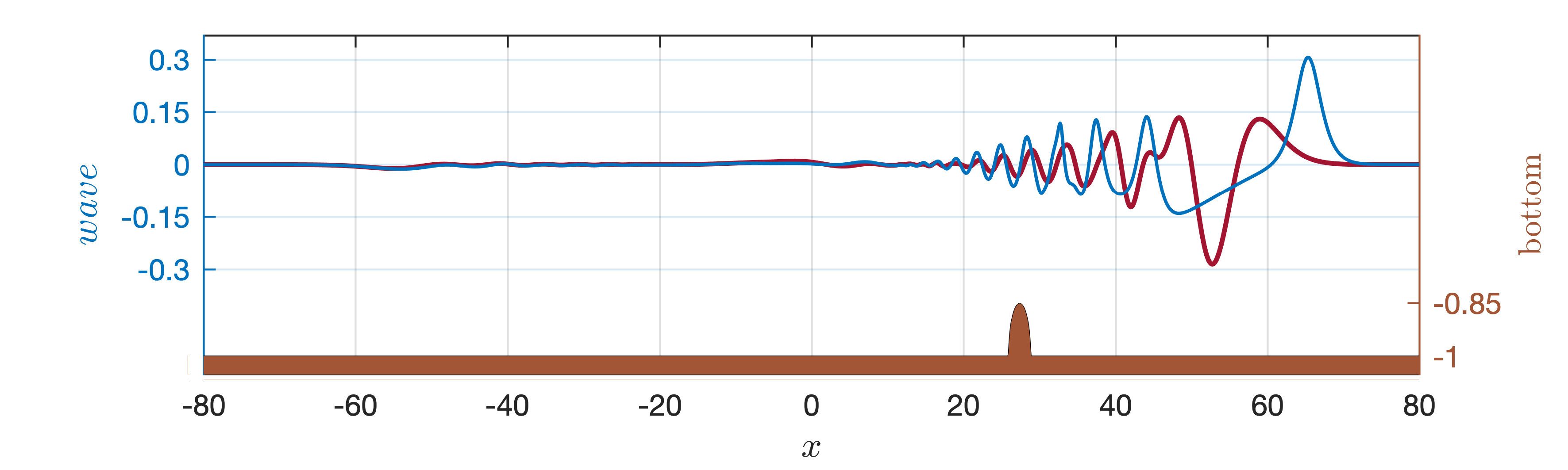}
        \caption{Generated waves considering $Fr = 0.125$ (top) and $Fr = 1.2$ (bottom), both for $t = 60$. Blue line: nonlinear model; red line: linear model.}
        \label{generated_waves_different_Froude_numbers}
\end{figure}
\begin{figure}[!htb]
         \centering
          \includegraphics[scale=1]{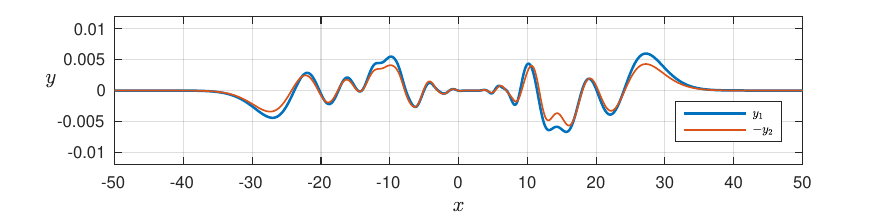}
          \includegraphics[scale=1]{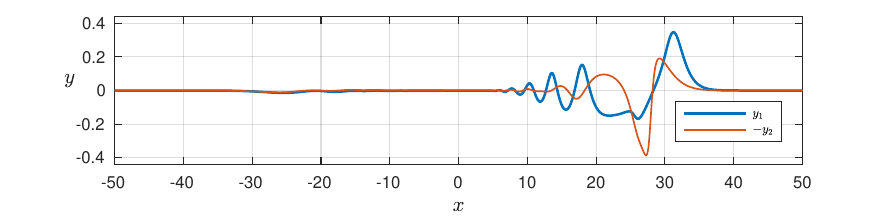}
        \caption{Symmetry comparison between waves generated for Froude numbers $Fr = 0.125$ (top) and $F = 1.2$ (bottom), for $t = 25$. Blue line: wave generated for $A > 0$; orange line: the opposite of wave generated for $A < 0$.}
        \label{figure_symmetry_comparison}
\end{figure}

\subsubsection{Froude number effect (Wave profile and symmetry)}

As shown in Figure \ref{generated_waves_different_Froude_numbers}, nonlinear effects are more pronounced as the Froude number increases. The nonlinear wave presents a leading crest followed by an oscillatory tail. The solution further shows that the wave trains radiated in the onshore and offshore directions differ substantially at a higher Froude number. The observed asymmetry was not reported explicitly by Chen et al. \cite{Chen:2024}, whose study neglected the acceleration stage of the bottom motion and considered Froude numbers only up to $Fr = 0.8$, although the authors did observe appreciable differences between the free-surface profiles obtained for $Fr = 0.4$ and $Fr = 0.8$.


The location of the maximum surface elevation follows a non-monotonic pattern in the Froude number. As noted by Zhao et al. \cite{Zhao:2026}, at low Froude numbers, the maximum wave amplitude is attained at the leading crest, a property that is already lost at $Fr \approx 0.5$, as Figure~\ref{benchmark_waves_Whit_17_case3} illustrates. Our results indicate that the trend reappears at $Fr > 1$: as shown in Figure~\ref{generated_waves_different_Froude_numbers}, the maximum amplitude is again located at the front crest.

An accurate prediction of the leading crest is essential for assessing wave impact in a tsunami context, and it is precisely this quantity that linear theory underestimates. We also note that at high Froude number the profile develops the N-wave shape commonly observed in earthquake-generated tsunamis \cite{Tadepalli:1994}.

Another important difference between the nonlinear and linear models is that the former is not symmetric with respect to variations in the topography amplitude. In other words, if we consider two experiments that differ only in the sign of the amplitude (but have the same magnitude), the linear model produces two symmetric solutions: the second is simply the reflection of the first about the $x$-axis. However, this symmetry does not hold for our nonlinear solution. As shown in Figure \ref{figure_symmetry_comparison}, the wave generated by a negative amplitude is not the mirrored version of the wave corresponding to a positive amplitude. We set $t = 25$ since the wave generated for the higher Froude number and $A<0$ depicts a sharper leading crest, which indicates wave breaking. Summarizing, at higher Froude numbers, nonlinearity yields asymmetry of the wave profile.


\subsubsection{Influence of the topography shape}

So far our computations have been following the exact experimental setup of Whittaker et al. \cite{Whittaker:2015,Whittaker:2017}, so only a semi-elliptical topography has been used. Motivated by the studies presented in \cite{Lo:2017}, we consider seven different geometric shapes for the topography: a Gaussian curve, a parabolic cap, a quartic cap, a rectangle, and three types of triangular profiles, all keeping the same area, height, and center of mass. 

Our results, based on potential theory, ignore the effects of vorticity and viscosity in the flow. This is in contrast with the studies reported in the literature, see \cite{Jing:2020_2,Lo:2017,Lo:2021,Sulvianuri:2025}, which are based on the Navier Stokes equations for the movement of a rigid landslide body along a slope \cite{Wang:2022}. 

\begin{figure}[htb]
    \centering
     \includegraphics[scale=1]{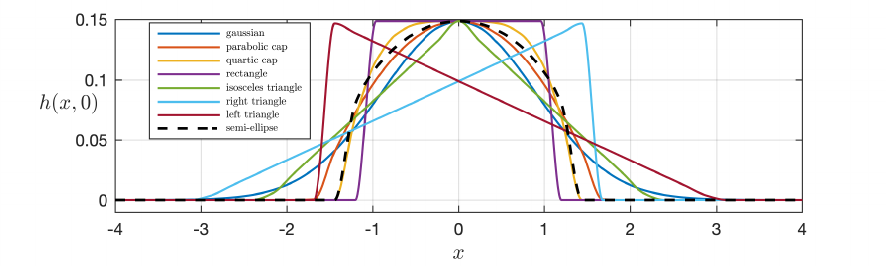}
     \includegraphics[scale=1]{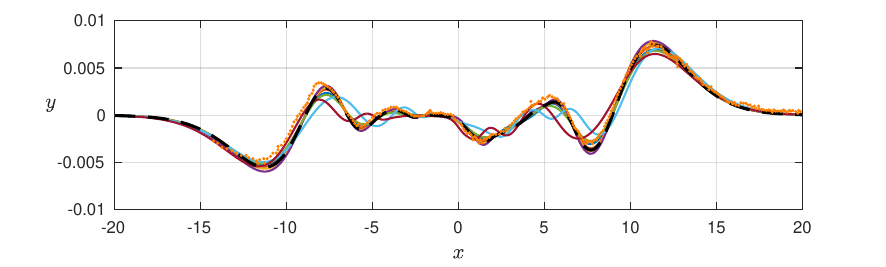}
    \caption{Topographies and wave generated using Hermite interpolation for different shapes of topography, while keeping the area constant. The Froude number is set to $Fr = 0.125$, the time shown is $t = 13.5$ and the other parameters are the same as the previous section. The orange dots in the bottom graph represent the experimental data. }
    \label{different_shapes_same_area}
\end{figure}

All the different topographies with their the corresponding wave profiles generated at time $t = 13.5$ are presented in Figure \ref{different_shapes_same_area}.

\begin{figure}[!htb]
    \centering
     \includegraphics[scale=1]{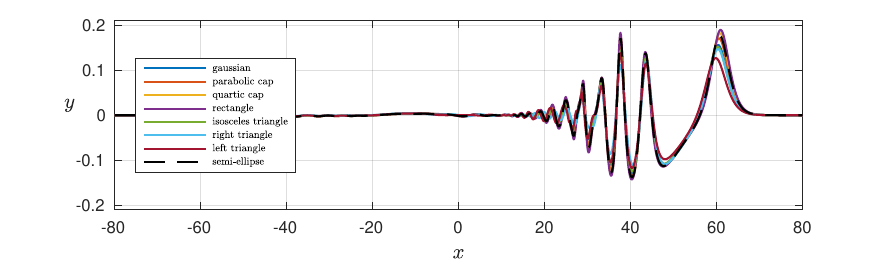}
    \caption{Topographies and wave generated using Hermite interpolation for different shapes of topography, while keeping the area constant. The Froude number is set to $Fr = 1.2$ and the time shown is $t = 60$.
    }
    \label{different_shapes_same_area_high_Froude}
\end{figure}

We also consider the high-Froude number case at time $t = 60$, which led to the same conclusion: the wave profile is kept essentially the same, with some small differences in the amplitude and in the oscillatory tail. See Figure \ref{different_shapes_same_area_high_Froude}.

These results indicate that the specific shape of the topography is not of significant importance, which supports the claims of Lo and Liu findings \cite{Lo:2017,Lo:2021} for subcritical regimes. 


\subsubsection{Topography amplitude and acceleration}

\begin{figure}[!htb]
         \centering
          \includegraphics[scale=1]{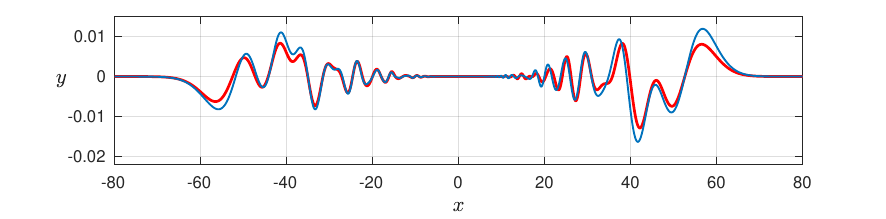}
        \caption{Generated waves considering $Fr = 0.125$, $A = 0.2972$ and $t = 60$. Blue line: nonlinear model; red line: linear model.}
        \label{generated_waves_different_amplitudes}
\end{figure}

Next we look at the effect of increasing the amplitude of the moving topography on the resulting wave. We keep the same parameters used before with  $Fr = 0.125$ as in Figure \ref{generated_waves_different_Froude_numbers}, and consider an amplitude twice bigger, $A = 0.2972$.
We observe that the nonlinear model behaves similarly to the linear one (see Figure \ref{generated_waves_different_amplitudes}). 
The amplitude of the main peak is nevertheless slightly higher in the nonlinear model, in agreement with the observations of Chen et al. \cite{Chen:2024} for topographic amplitudes up to $0.15$. Doubling the amplitude therefore does not lead to a departure from linearity as the increase of the Froude number does, confirming the role of the Froude number as the dominant parameter.

\begin{figure}[htb]
         \centering
          \includegraphics[scale=1]{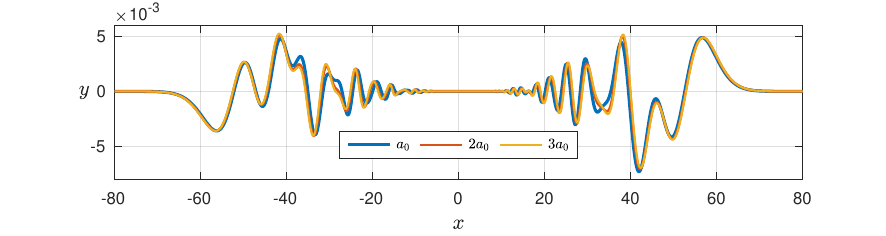}
         \includegraphics[scale=1]{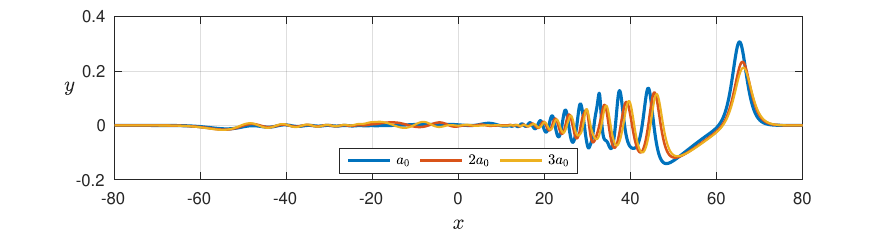}
        \caption{Generated waves considering different values of acceleration for $t = 60$, with $Fr = 0.125$ (top) and $Fr = 1.2$ (bottom).}
        \label{generated_waves_different_accelerations}
\end{figure}

Now we focus on the acceleration effect. Following Whittaker et al. \cite{Whittaker:2015}, they reported that the initial acceleration is less important than the terminal velocity, for $Fr = 0.375$, than it is for $Fr = 0.125$. We investigate this feature with our numerical scheme, considering $a_0 = 0.153$ as the reference value and work with three different accelerations: $a_0$, $2a_0$, and $3a_0$. The terminal velocity of each three cases is kept at $Fr = 0.125$ and  $Fr = 1.2$. For the motion with higher acceleration, the block reaches the constant velocity stage earlier. In addition, the duration of the constant velocity motion is the same for the three cases. The results are shown in Figure \ref{generated_waves_different_accelerations}. 

Changing the acceleration produces no significant change in the wave generated at low Froude number. At $Fr = 1.2$, however, an unexpected feature emerges: the largest leading wave is the one associated with the {smallest} acceleration. This reversal is not the trend usually reported in the literature \cite{Jing:2020,Whittaker:2017}. Whittaker et al. \cite{Whittaker:2017} investigated motions with Froude numbers up to $0.75$ and observed that the nonlinearity of the trailing waves tends to decrease at $Fr = 0.5$ for high accelerations ($a_0 \approx 2$), which is consistent with what we find at moderate Froude numbers. The profiles obtained with $2a_0$ and $3a_0$ are practically indistinguishable at high Froude number, which supports this reading. 

\subsection{Exploring nonlinear effects on the velocity field}

\begin{figure}[!htb]
         \centering
         \includegraphics[scale=1]{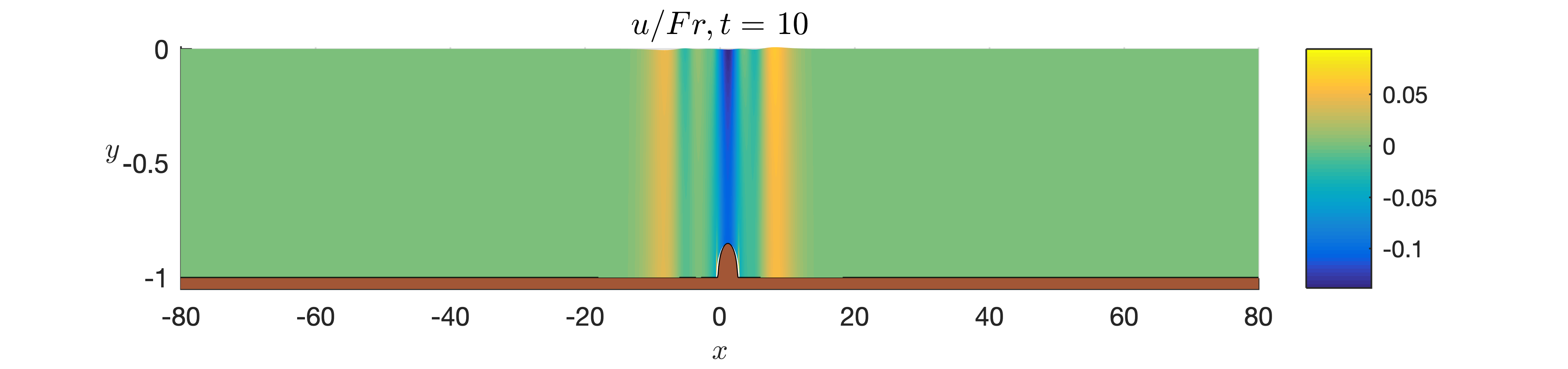}
         \includegraphics[scale=1]{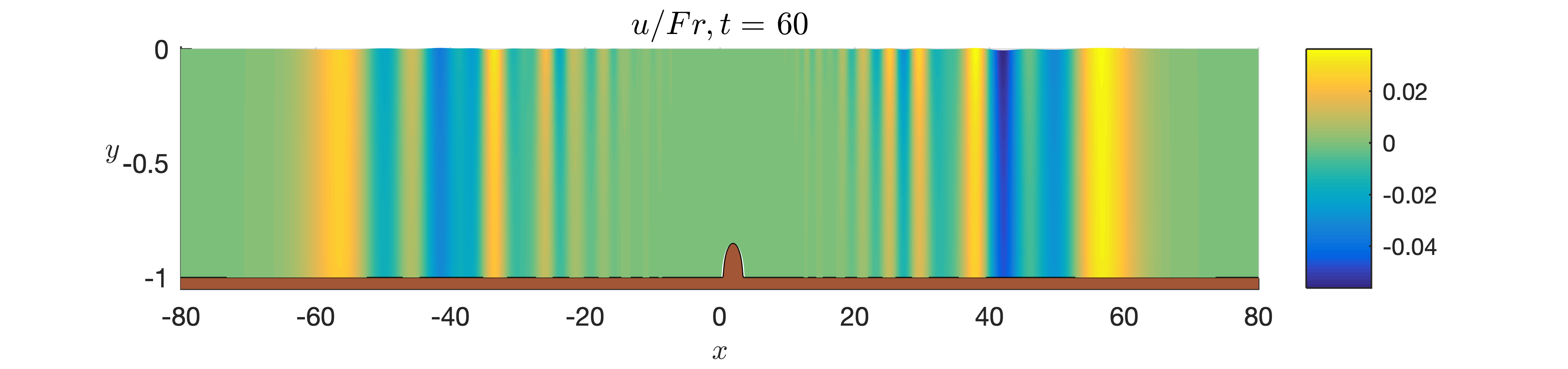}
        \caption{Horizontal normalized components of the velocity field, considering $Fr = 0.125$ for $t = 10$ and $t = 60$.}
        \label{velocity_field_Fr_0.125_u}
\end{figure}

\begin{figure}[!htb]
         \centering
         \includegraphics[scale=1]{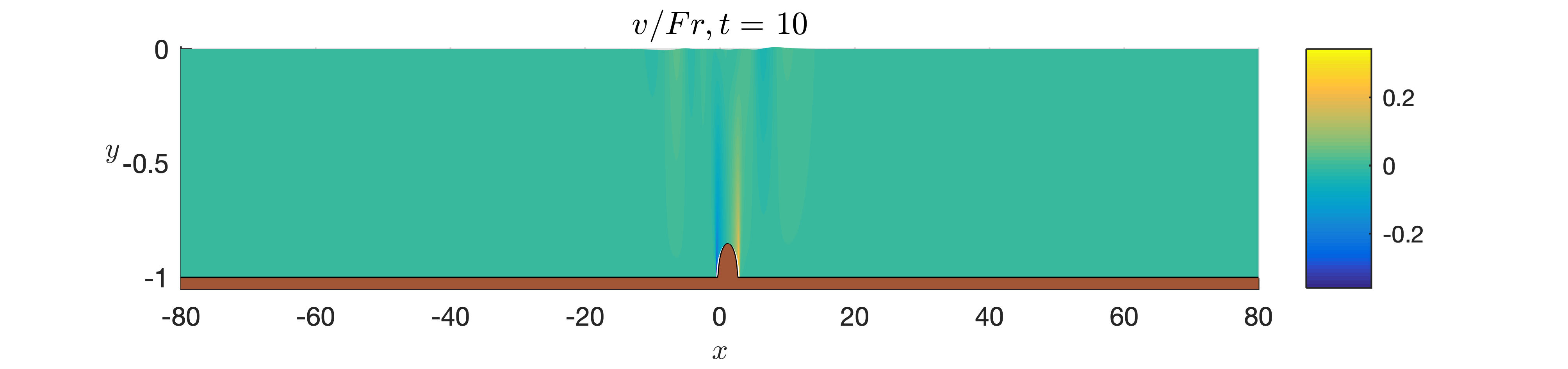}
         \includegraphics[scale=1]{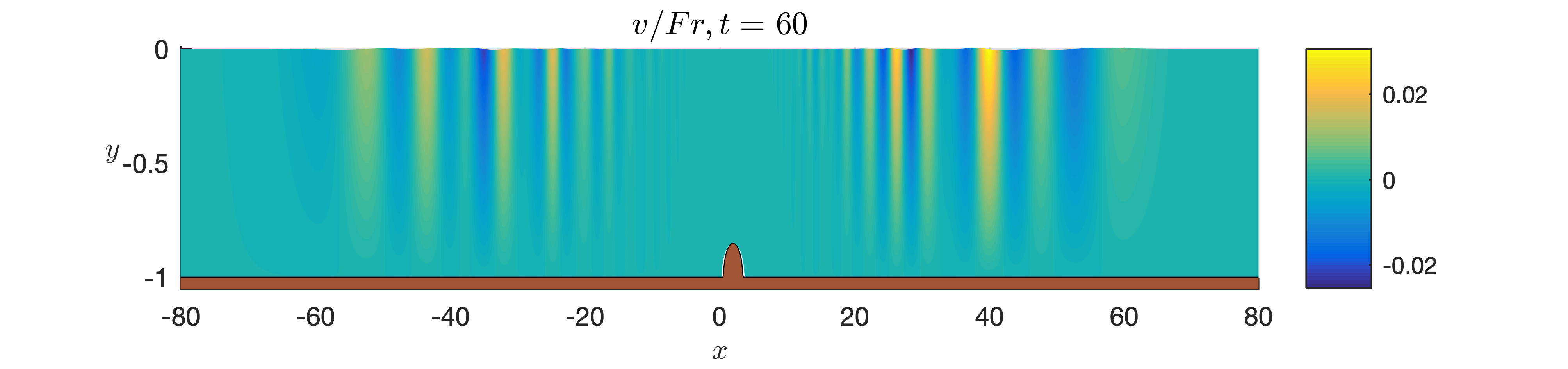}
        \caption{Vertical normalized components of the velocity field, considering $Fr = 0.125$ for $t = 10$ and $t = 60$.}
        \label{velocity_field_Fr_0.125_v}
\end{figure}

\begin{figure}[!htb]
         \centering
         \includegraphics[scale=1]{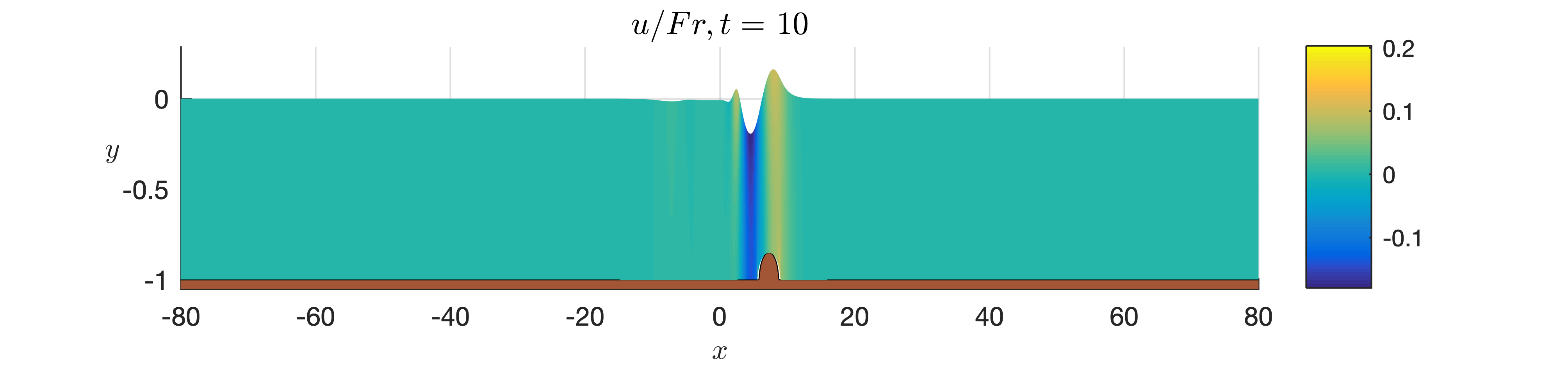}
         \includegraphics[scale=1]{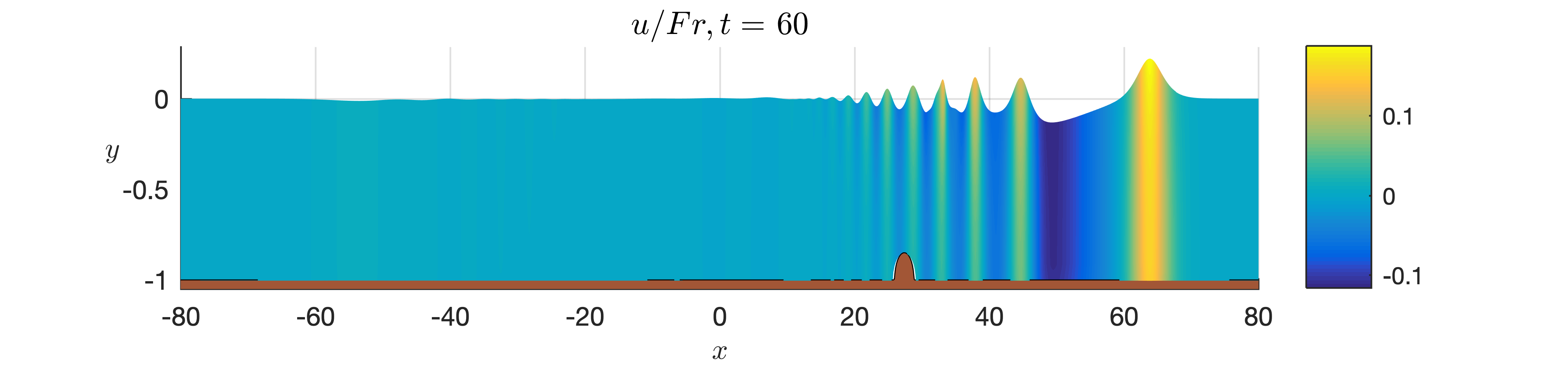}
        \caption{Horizontal normalized components of the velocity field, considering $Fr = 1.2$ for $t = 10$ and $t = 60$.}
        \label{velocity_field_Fr_high_u}
\end{figure}

\begin{figure}[!htb]
         \centering
         \includegraphics[scale=1]{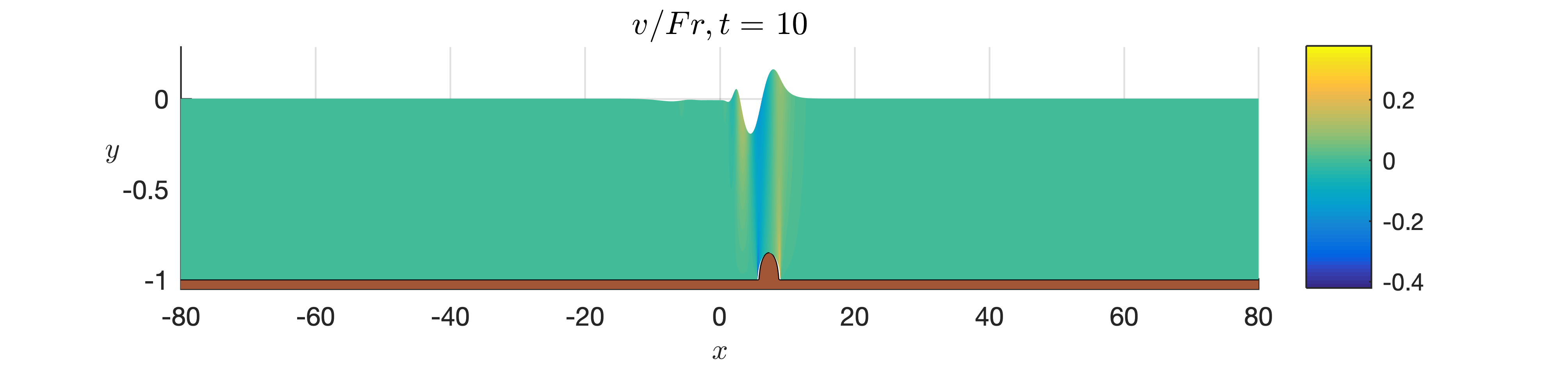}
         \includegraphics[scale=1]{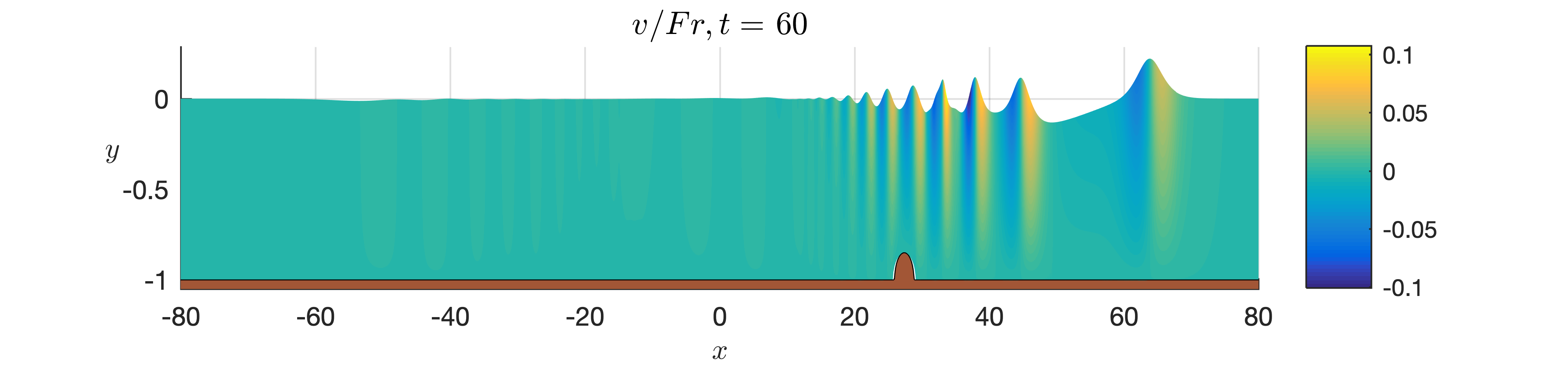}
        \caption{Vertical normalized components of the velocity field, considering $Fr = 1.2$ for $t = 10$ and $t = 60$.}
        \label{velocity_field_Fr_high_v}
\end{figure}

With the same setup for the waves as in Figure \ref{generated_waves_different_Froude_numbers}, we explore the velocity field beneath them, using equations \eqref{phi_x} and \eqref{phi_y}. The results are displayed, respectively, in Figures \ref{velocity_field_Fr_0.125_u} and \ref{velocity_field_Fr_0.125_v} for the low Froude number case and in Figures \ref{velocity_field_Fr_high_u} and \ref{velocity_field_Fr_high_v} for the high Froude number. 

\begin{figure}[!htb]
        \centering
        \includegraphics[scale=1]{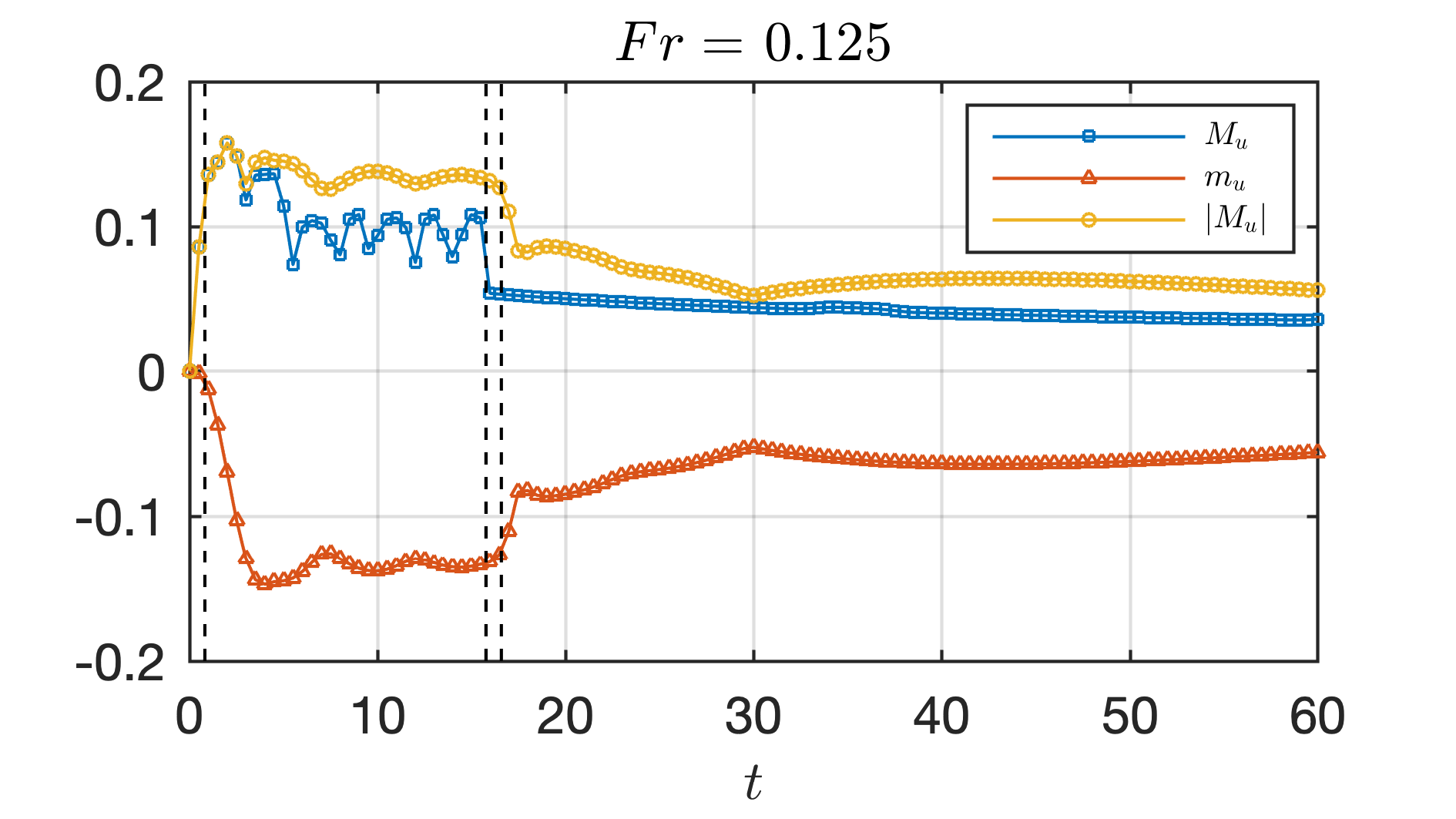}
        \includegraphics[scale=1]{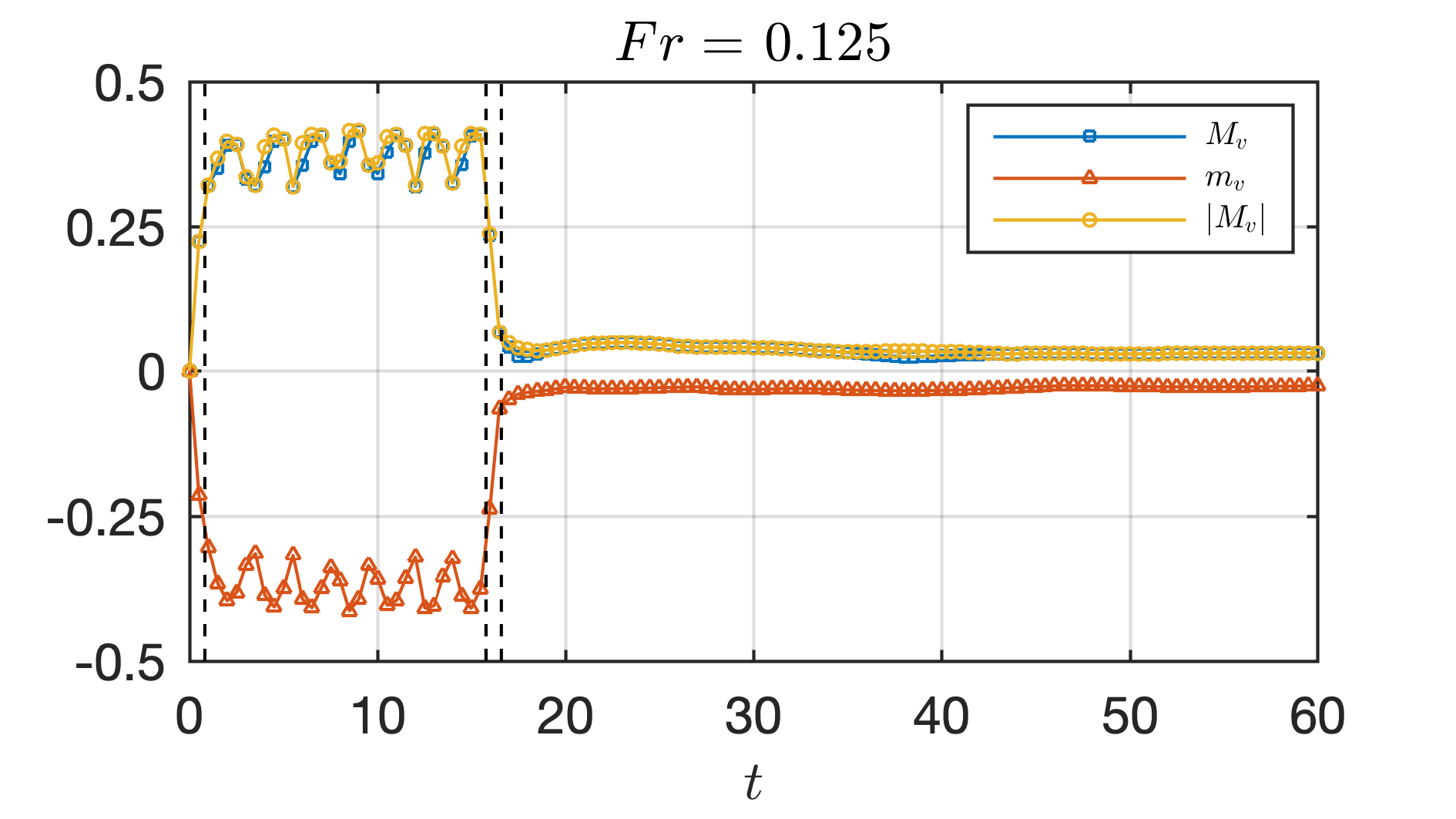}
        \includegraphics[scale=1]{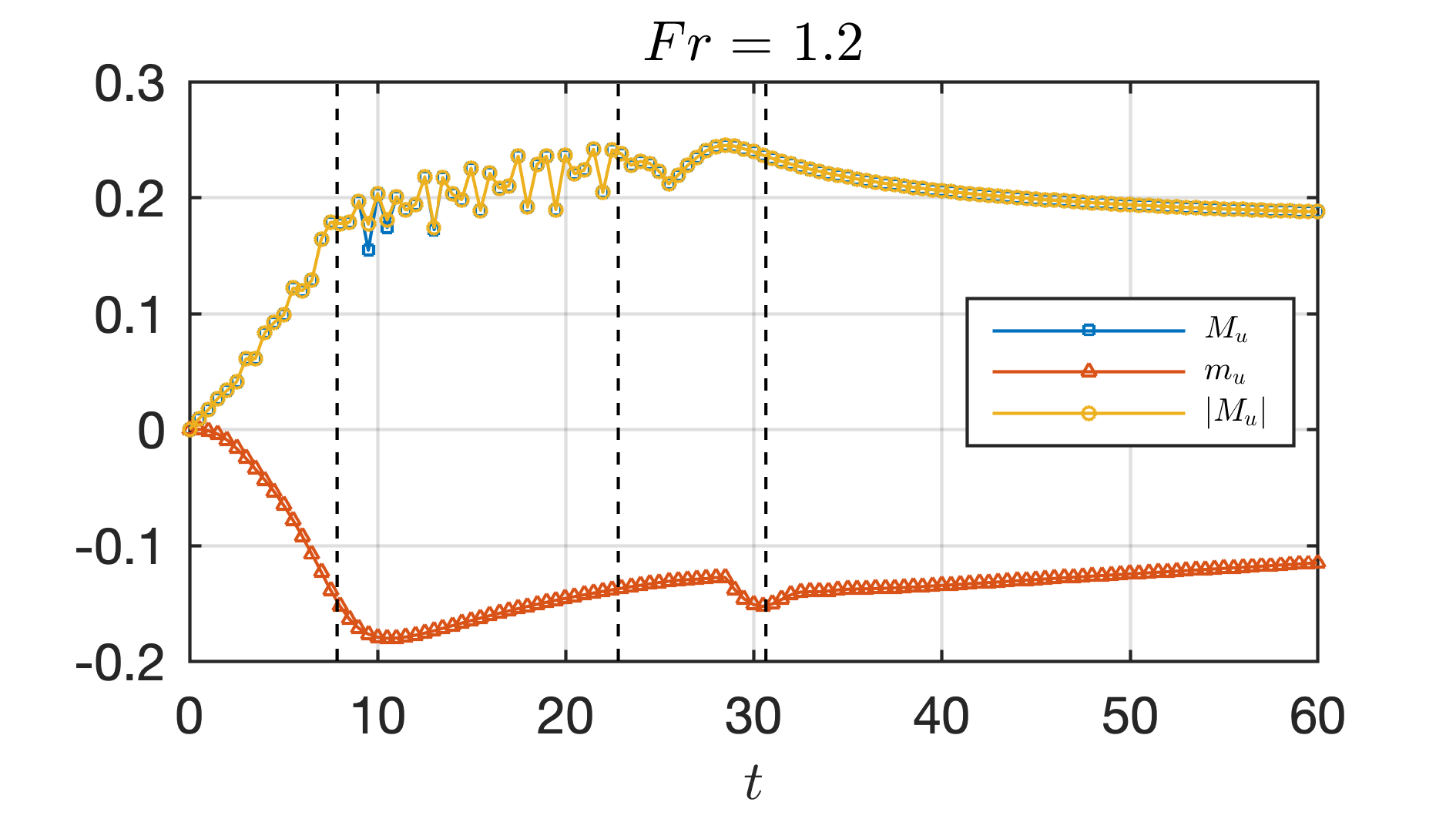}
        \includegraphics[scale=1]{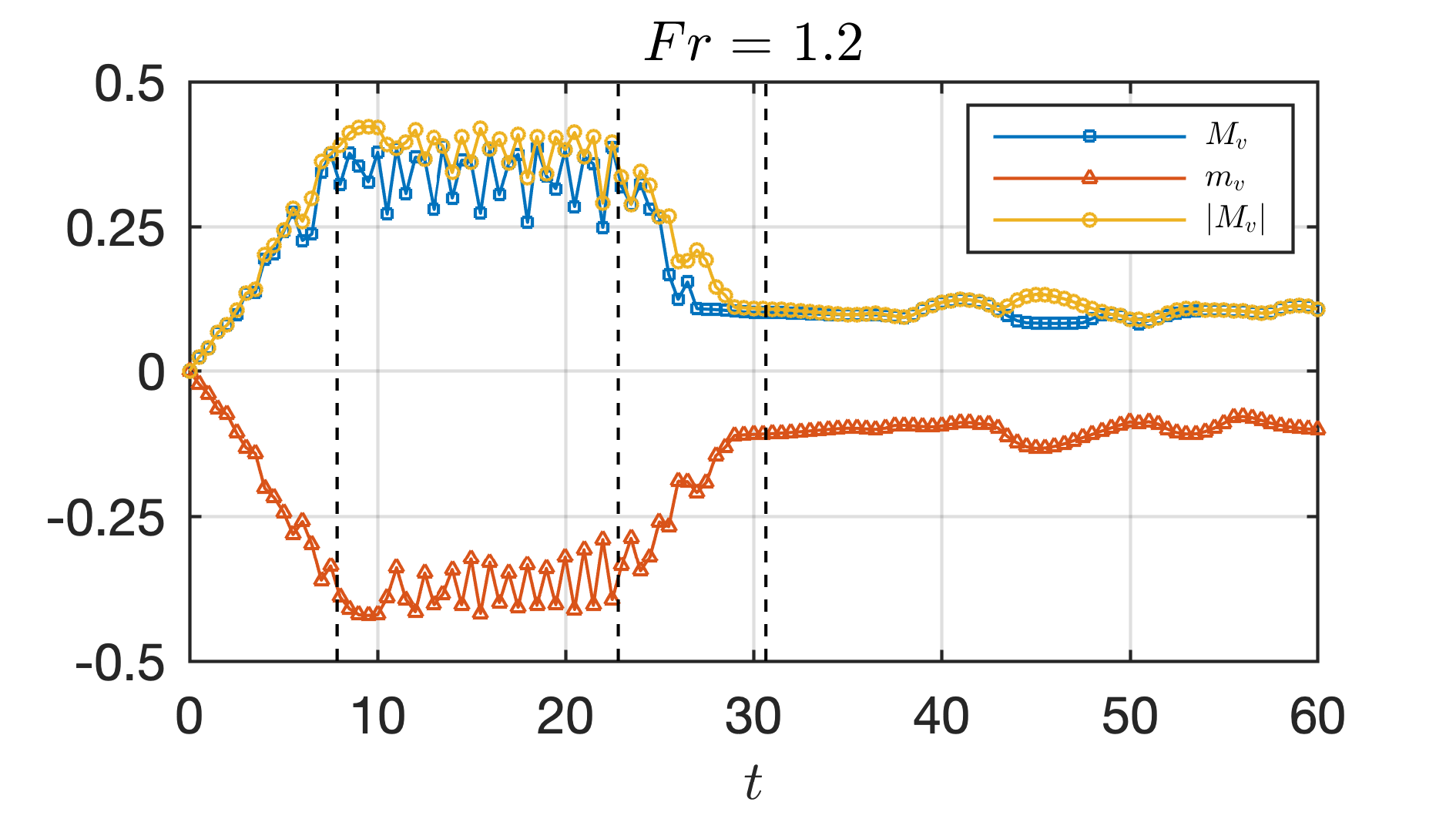}
        \caption{Horizontal and vertical components of the velocity field, considering $Fr = 0.125$ (top) and $Fr = 1.2$ (bottom). Vertical dotted lines indicate the transition times between the different stages of the block motion: uniform acceleration from rest, motion at constant velocity, and uniform deceleration until the block comes to rest.}
        \label{velocity_field_maximum_history}
\end{figure}

We chose these two times since, at $t = 10$ the block is in its phase of maximum velocity for both cases and at $t = 60$ the wave is already in its propagation phase after the block has fully stopped. These differences are reflected in the extrema of the velocity field, as shown in Figure \ref{velocity_field_maximum_history}. Specifically, Figure \ref{velocity_field_maximum_history} plots the maximum and minimum values of the normalized horizontal and vertical velocity components, together with their maximum absolute values. We define
\[
M_u(t)=\max_{(x,y)\in\Omega}\frac{u(x,y,t)}{Fr},\qquad
m_u(t)=\min_{(x,y)\in\Omega}\frac{u(x,y,t)}{Fr},
\]
and
\[
|M_u(t)|=\max_{(x,y)\in\Omega}\frac{|u(x,y,t)|}{Fr}.
\]
Similarly, we define $M_v(t)$, $m_v(t)$, and $|M_v(t)|$ for the vertical velocity component $v$.

We observe that the vertical component of velocity shows maximum values that are greater than the horizontal component, in both cases, for both instants of time, as can be checked in Table \ref{maximum_velocity_table}.
\begin{table}[!ht]
    \centering
    \begin{tabular}{|c|c|c|} \hline
        $Fr$ & $ \max\limits_{t \in [0,60]} |M_u(t)| $ & $ \max\limits_{t \in [0,60]} |M_v(t)| $  \\ \hline
        $0.125$ & $0.157$ & $0.414$ \\
        $1.2$ & $0.245$ & $0.422$ \\ \hline
    \end{tabular}
    \caption{Maximum of the absolute value of the velocity field horizontal and vertical components.}
    \label{maximum_velocity_table}
\end{table}
Note also that, once the block stops, the maximum of the vertical component decays much faster than that of the horizontal one, since the vertical velocity is driven directly by the bottom motion whereas the horizontal one remains associated with the propagating wave. 

The offshore direction is also markedly more energetic than the onshore one at high Froude number, as Figures \ref{velocity_field_Fr_high_u} and \ref{velocity_field_Fr_high_v} show, whereas at $Fr = 0.125$ the velocity field is nearly symmetric about the origin. 


\subsection{Particle dynamics}

\begin{figure}[htb]
         \centering
         \includegraphics[scale=1]{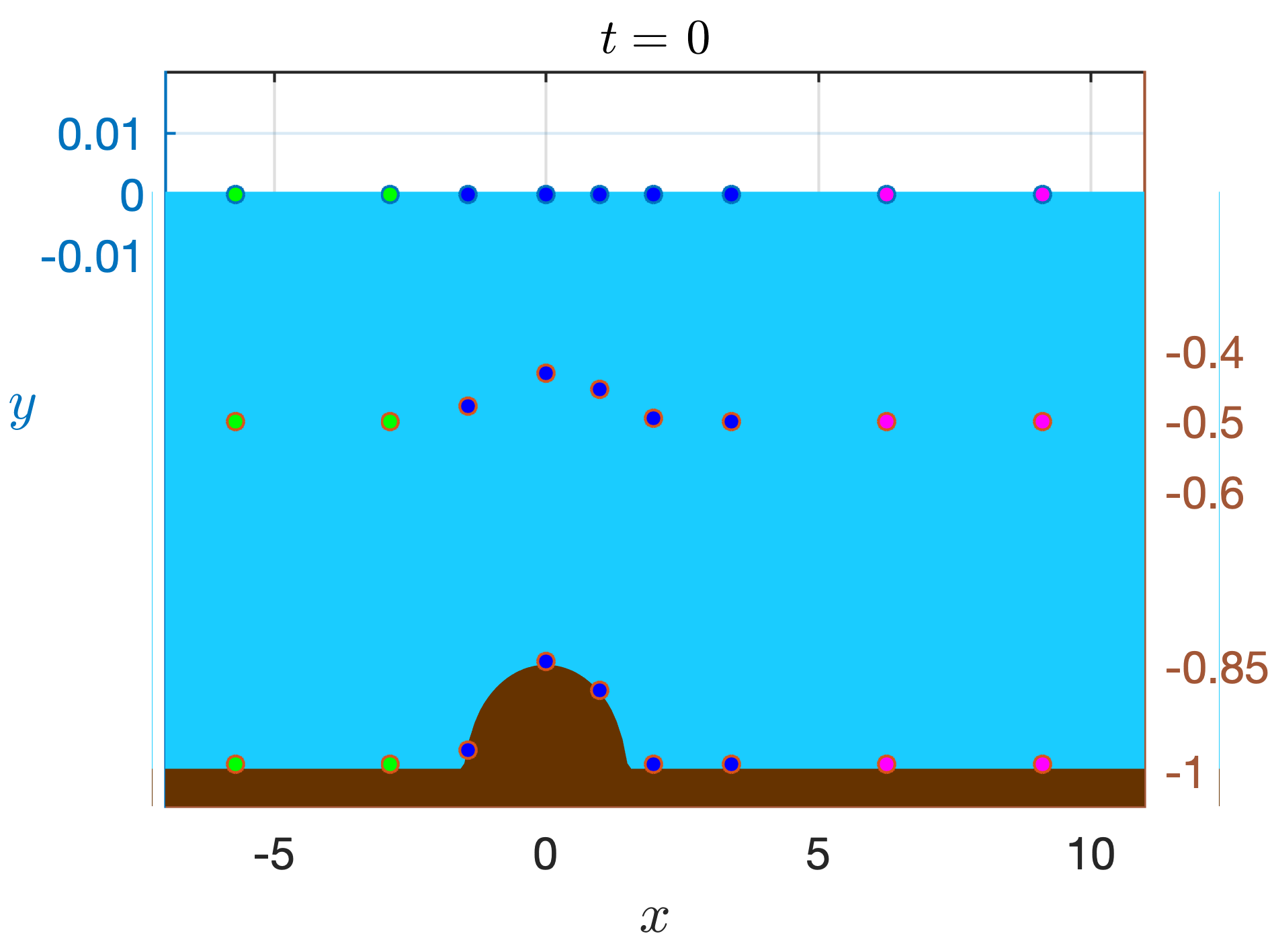}
         \includegraphics[scale=1]{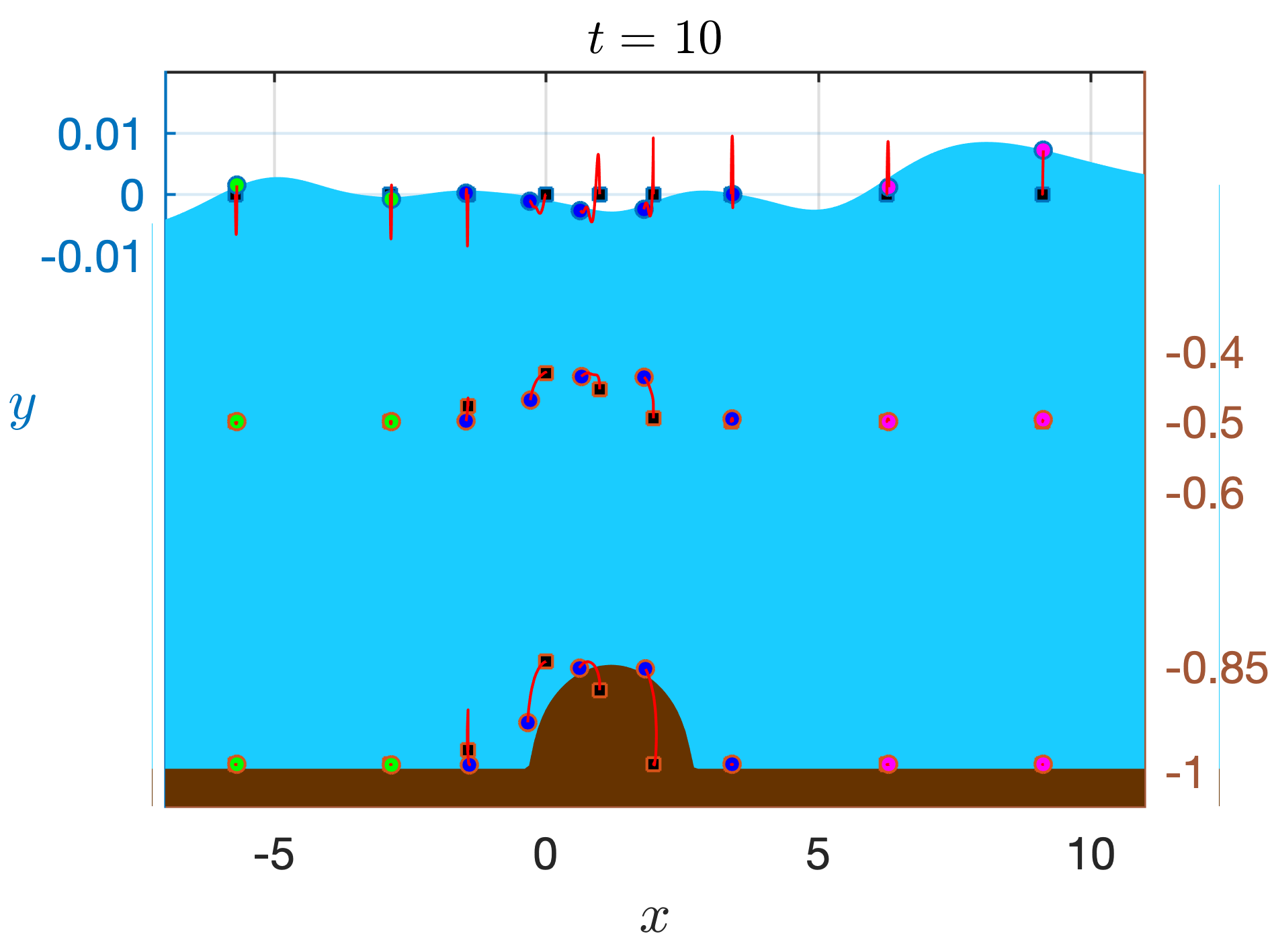}
         \includegraphics[scale=1]{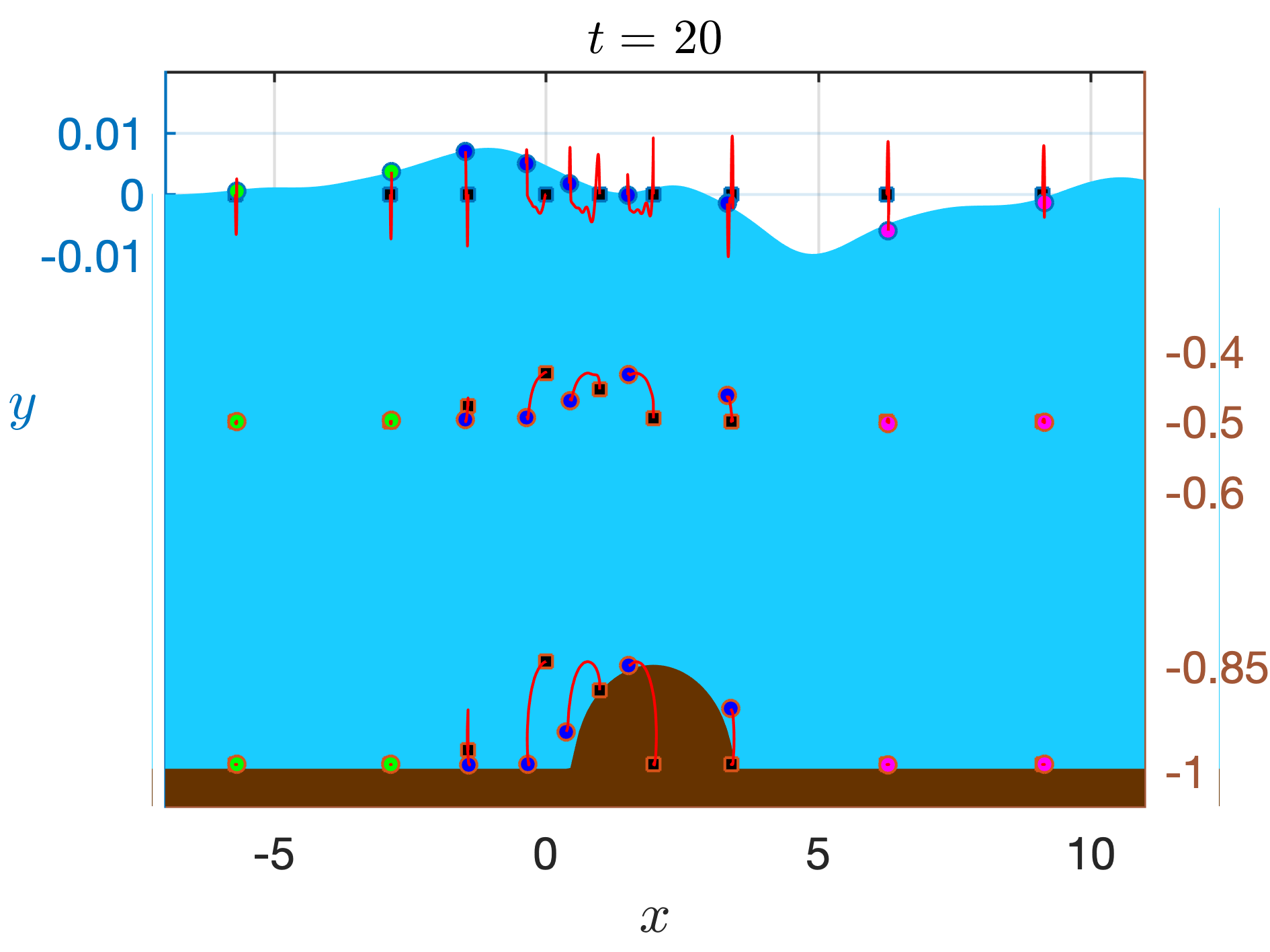}
         \includegraphics[scale=1]{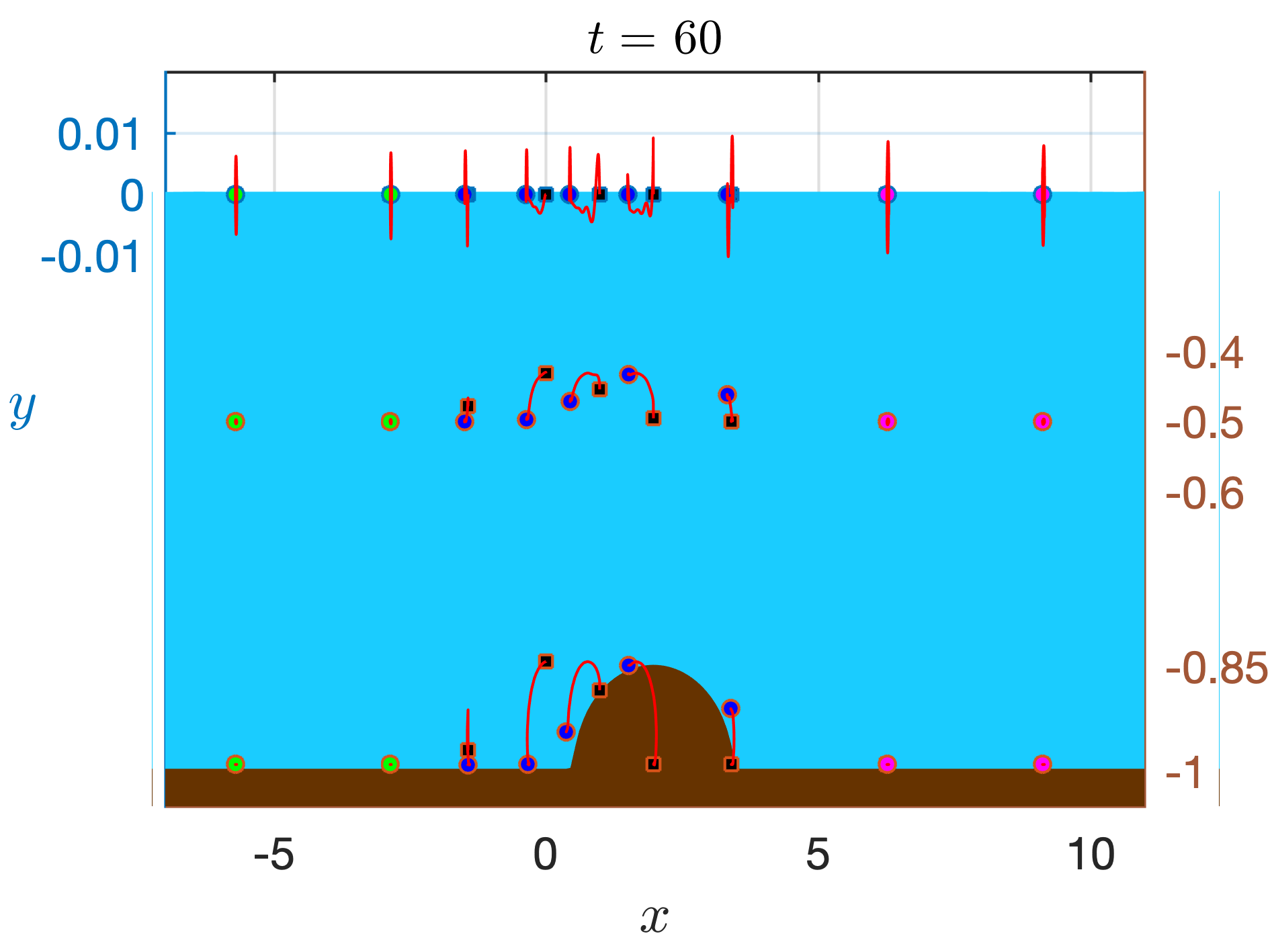}
        \caption{Particle dynamics history for $Fr = 0.125$. Colors green, blue and pink represent, respectively, particles on the onshore, main and offshore fluid regions. The black square indicates where the particles starts its motion.}
        \label{particle_snapshots_low_Fr}
\end{figure}

\begin{figure}[htb]
         \centering
         \includegraphics[scale=1]{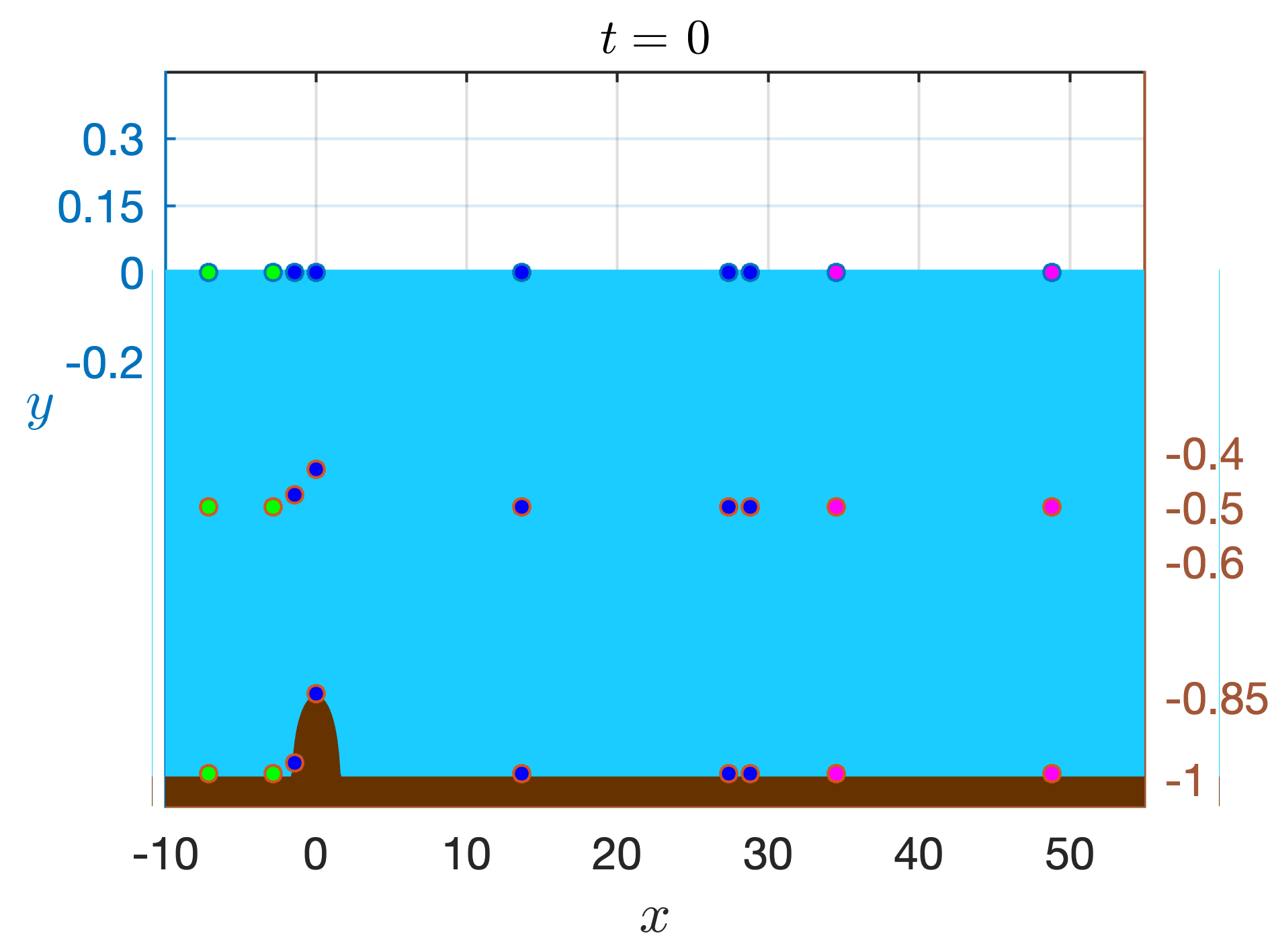}
         \includegraphics[scale=1]{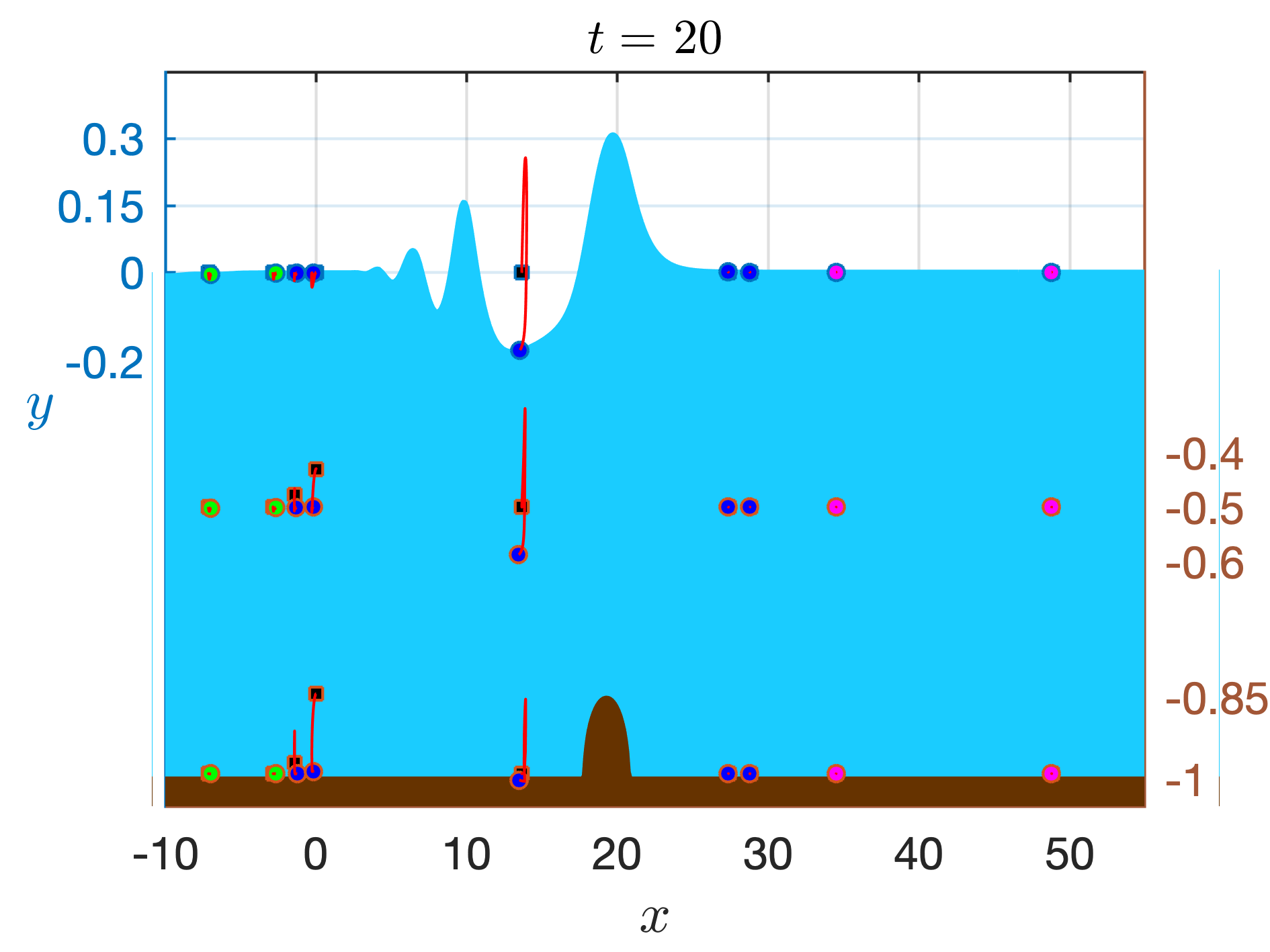}
         \includegraphics[scale=1]{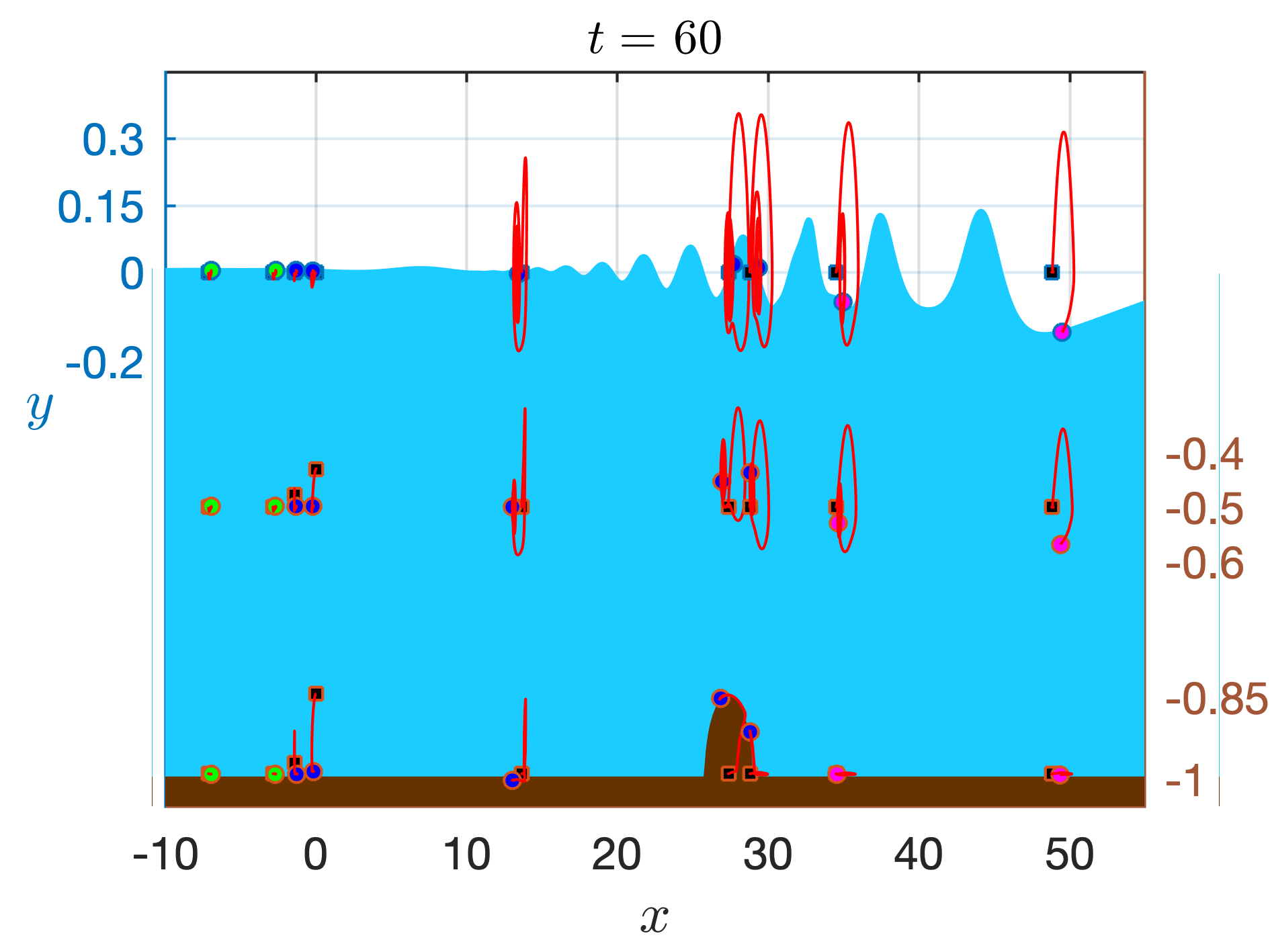}
         \includegraphics[scale=1]{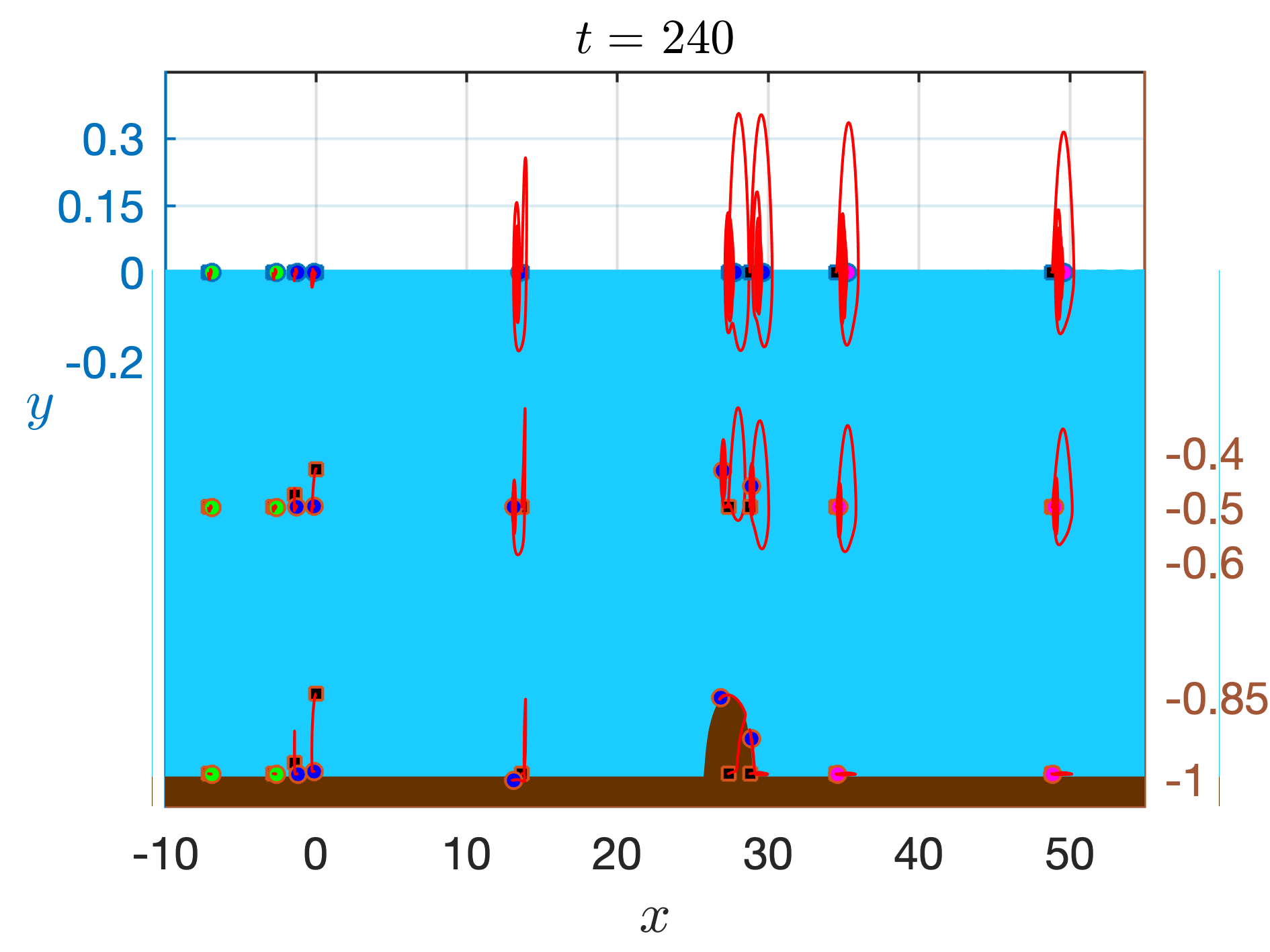}
        \caption{Particle dynamics history for $Fr = 1.2$. Colors green, blue and pink represent, respectively, particles on the onshore, main and offshore fluid regions. The black square indicates where the particles start their motion.}
        \label{particle_snapshots_high_Fr}
\end{figure}

We now integrate system \eqref{system_particle_canonical} for the two scenarios considered above, $Fr = 0.125$ and $Fr = 1.2$, and examine the resulting Lagrangian dynamics. Since the flow is non-autonomous, the trajectories cannot be read off from a phase portrait and must be obtained by direct integration, which the formulation of Section~\ref{Formulation} makes possible. Snapshots of both computations are shown in Figures \ref{particle_snapshots_low_Fr} and \ref{particle_snapshots_high_Fr}.  

We released 6 particles in the onshore region, 15 in the main region and 6 in the offshore region, distributed over three levels of the water column: the free surface, the seabed and mid-depth. The computation reproduces the two kinematic boundary conditions exactly: particles released on the free surface remain on it throughout the simulation, and particles released on the seabed remain on it at every instant, including while the obstacle passes underneath them. Since these constraints are not imposed on the trajectory solver but follow from formulas \eqref{Gamma_equation} and \eqref{Lambda_equation}, their fulfilment is a stringent test of the numerical scheme.

For the high Froude number case, we use a longer observation time, since the surface wave above the particles in the offshore region takes longer to leave the region and for the fluid to return to rest. We therefore employ a computational domain four times larger, so that the particles have time to come to rest before the periodic images of the wave re-enter the domain. Since, the generated wave is more pronounced, it causes the particles to undergo more orbital motions than in the previous cases. As a result, their net displacements are considerably larger. 

\begin{figure}[!htb]
         \centering
         \includegraphics[scale=1]{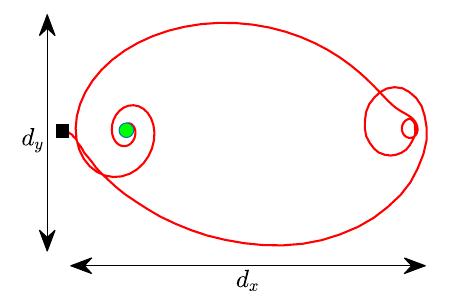}
         \includegraphics[scale=1]{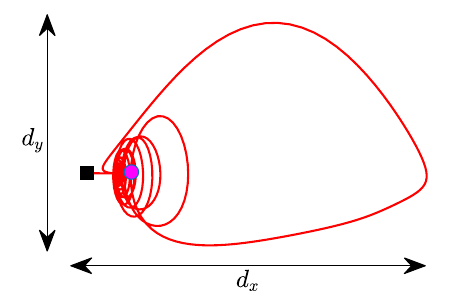}
        \caption{Zoomed-in view of the particle trajectories. The left panel shows the trajectory of a particle initially located at the free surface, with $x_0 \approx -6$ and $F_r = 0.125$ (see Figure \ref{particle_snapshots_low_Fr}). The right panel shows the trajectory of a particle initially located at mid-depth, with $x_0 \approx 50$ and $Fr = 1.2$ (see Figure \ref{particle_snapshots_high_Fr}). For the former, the horizontal and vertical displacements are $d_x \approx 0.0279$ and $d_y \approx 0.0128$, respectively, whereas for the latter, these values are $d_x \approx 1.3737$ and $d_y \approx 0.2134$. The black square and circle indicate the initial and final particle positions, respectively. 
        }
        \label{schematic_figure_particles_displacement}
\end{figure}

The study of particle dynamics has direct practical implications for submarine operations, pollutant dispersion and sediment transport \cite{Berchet:2018,Nachbin_RibeiroJr:2014}. To quantify the excursion of each particle we measure the difference between the extreme values attained along its path in the horizontal ($d_x$) and vertical ($d_y$) directions,
\begin{align*}
    d_x & = \max_{t \in [0,T]} x(t) - \min_{t \in [0,T]} x(t), \\
    d_y & = \max_{t \in [0,T]} y(t) - \min_{t \in [0,T]} y(t),
\end{align*}
where $T$ is the final time of the simulation. Figure \ref{schematic_figure_particles_displacement} illustrates both quantities on two trajectories taken from our computations. The left panel shows a particle trajectory located in the onshore region at the free surface for the low Froude number case. The dynamics is characterized by an initial loop on the right, corresponding to the first packet of waves propagating in the onshore direction. A second loop occurs when a second wave reaches the same location, as a consequence of the block deceleration.
On the other hand, the right panel shows a particle located in the offshore region, at mid-depth, for the high Froude number case. In this situation, when the wave reaches its position, it causes the particle to undergo several loops. In both cases, however, the particles exhibit a net displacement toward the right, i.e., in the direction of the block motion.

%


\begin{figure}[!htb]
         \centering
         \includegraphics[scale=1]{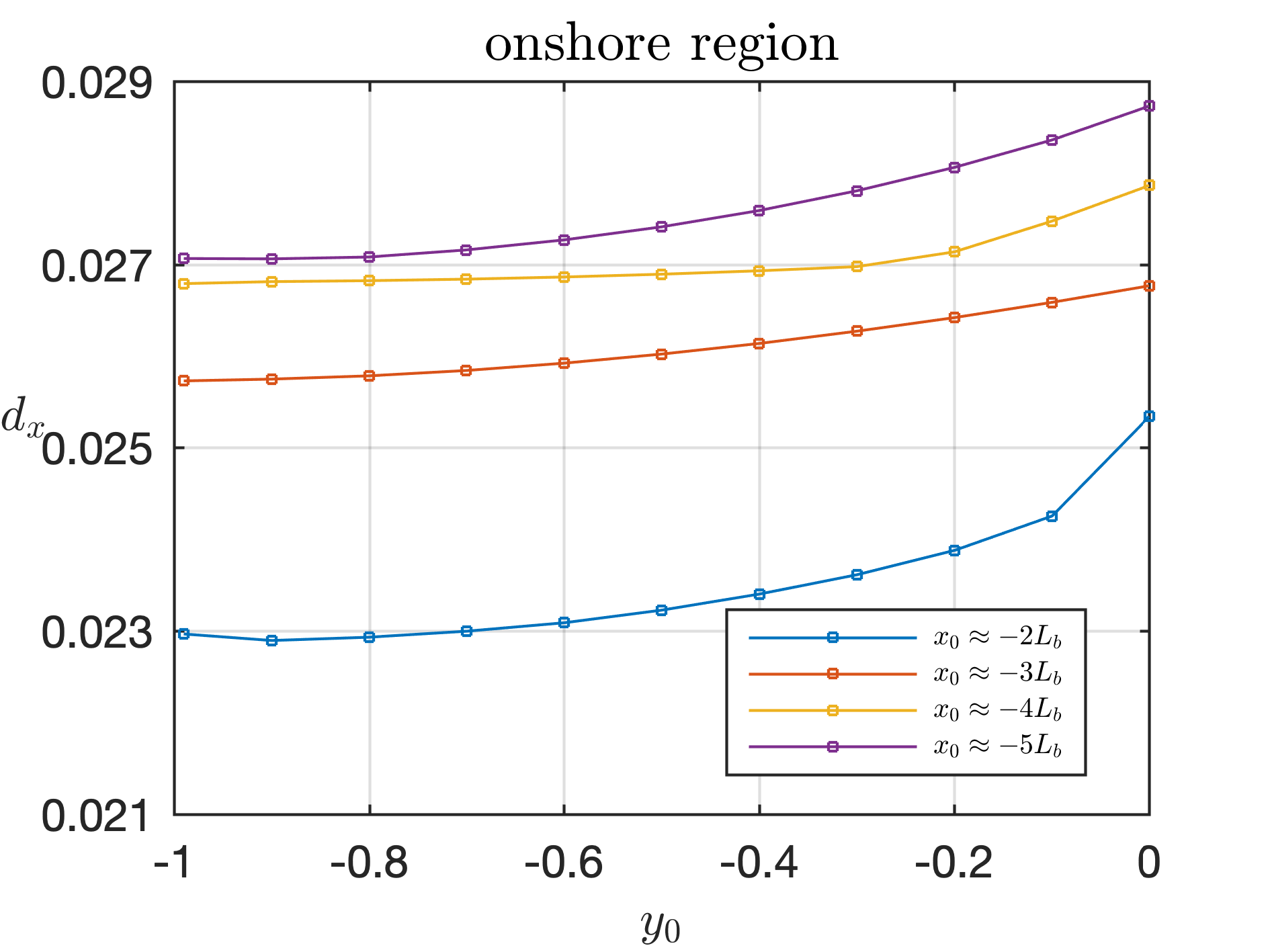}
         \includegraphics[scale=1]{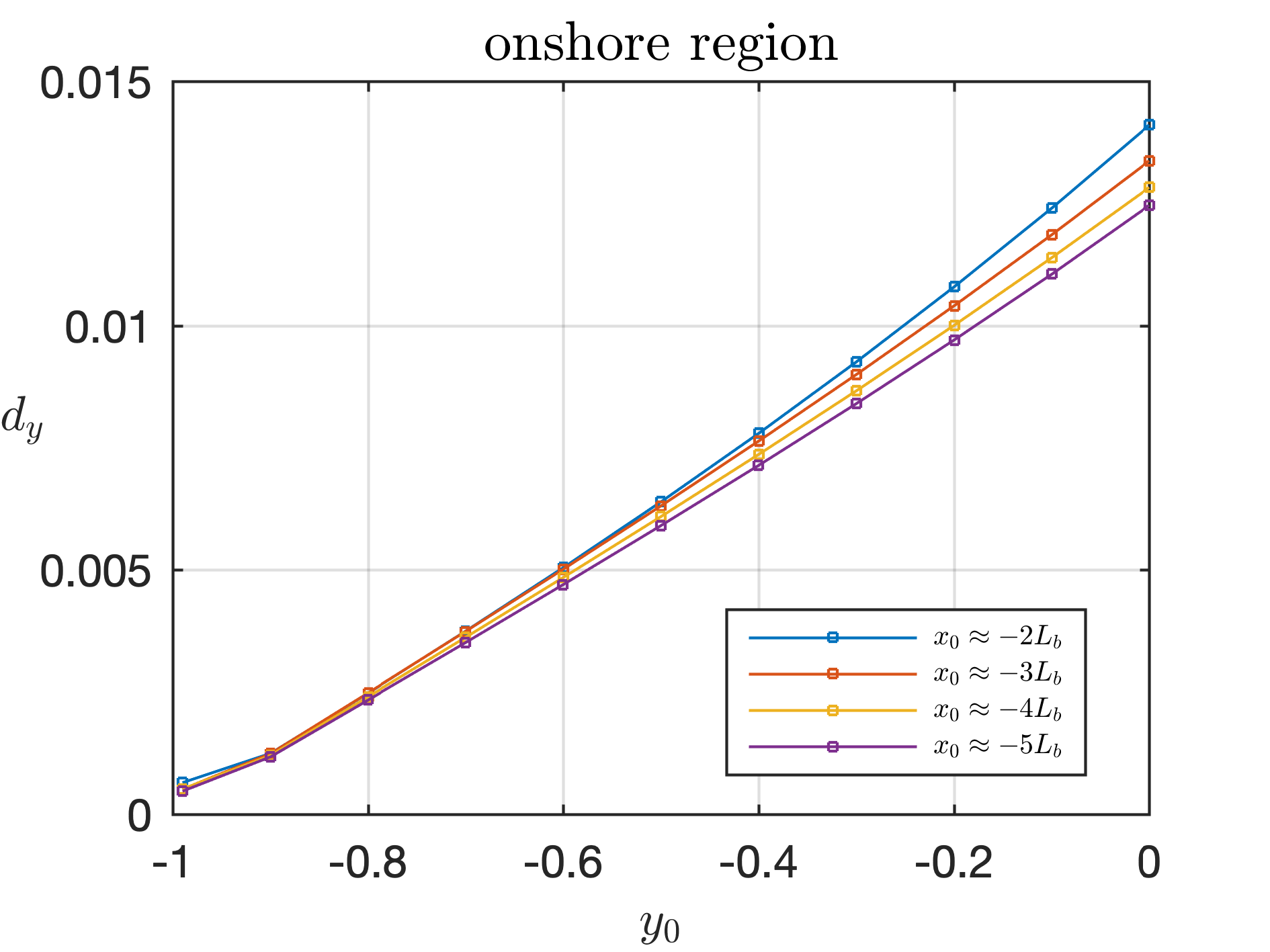}
         \includegraphics[scale=1]{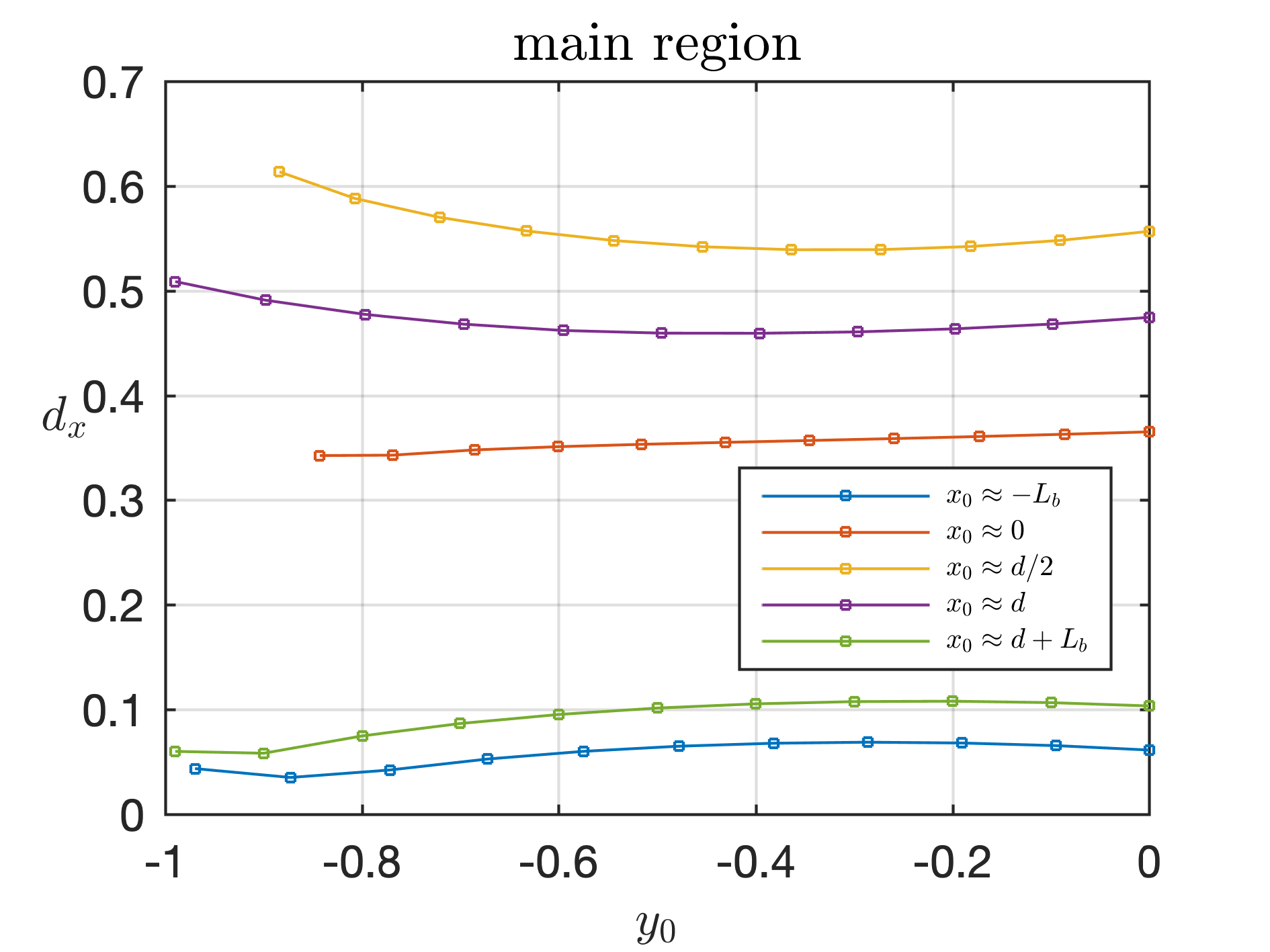}
         \includegraphics[scale=1]{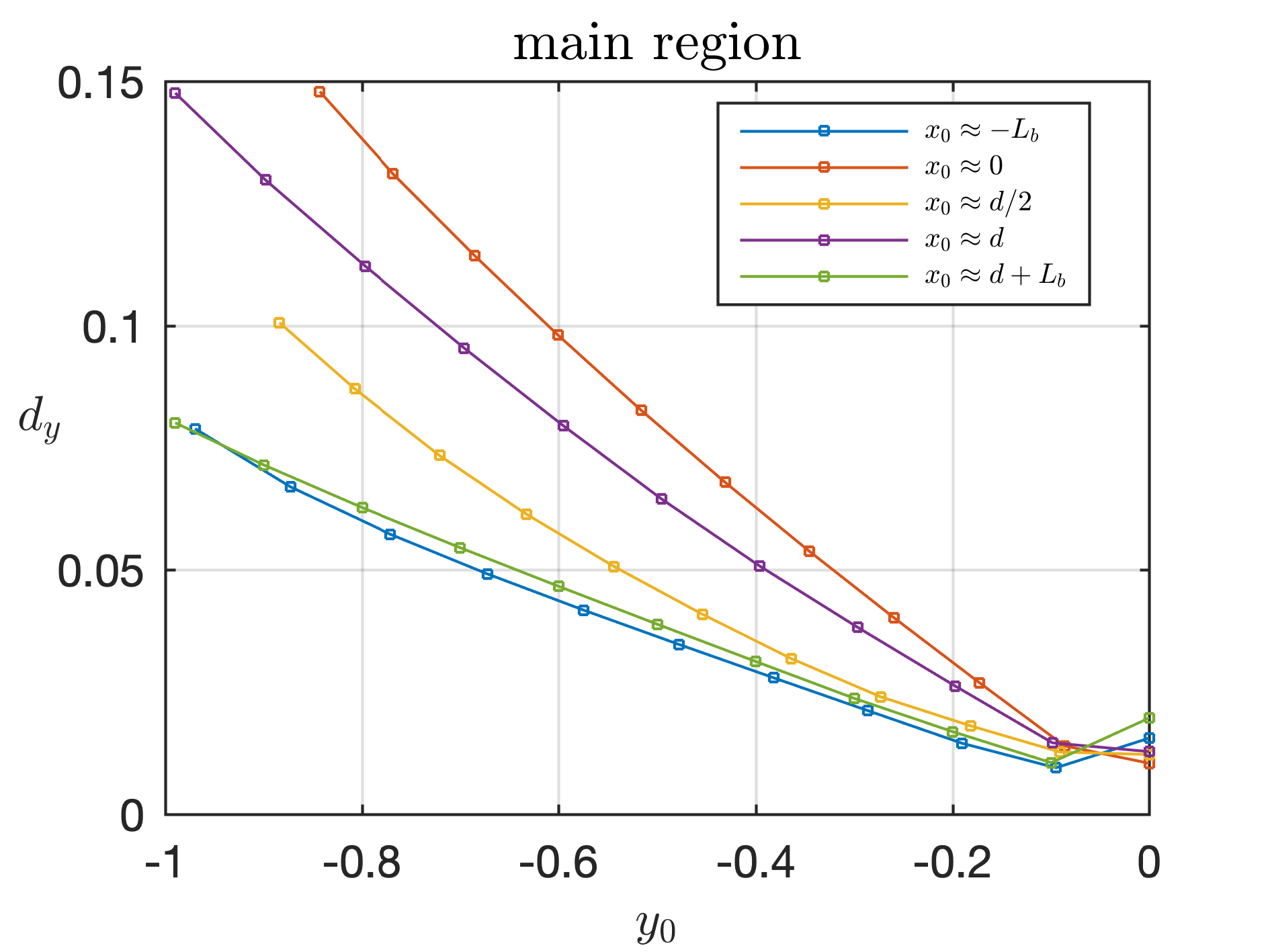}
         \includegraphics[scale=1]{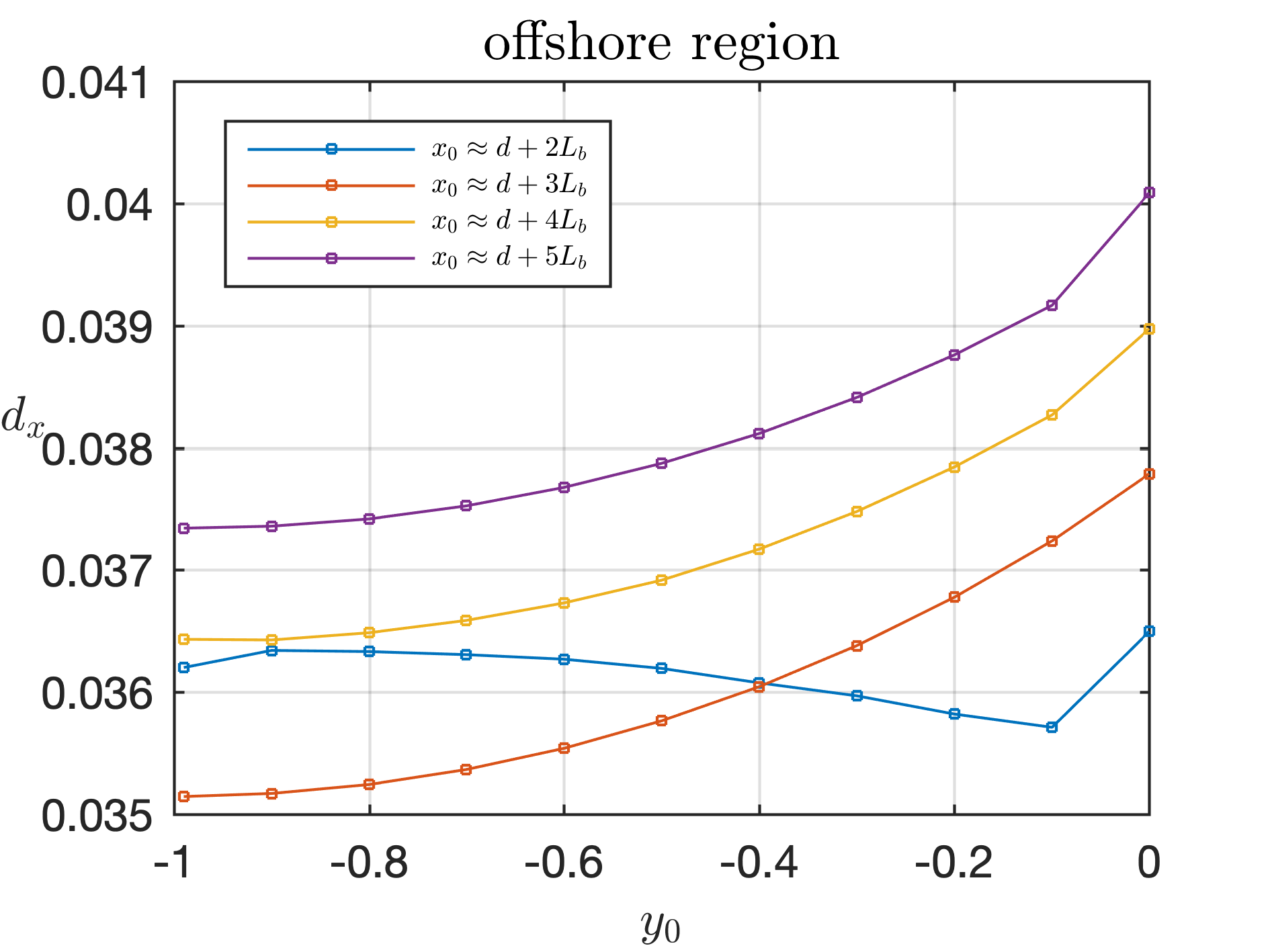}
         \includegraphics[scale=1]{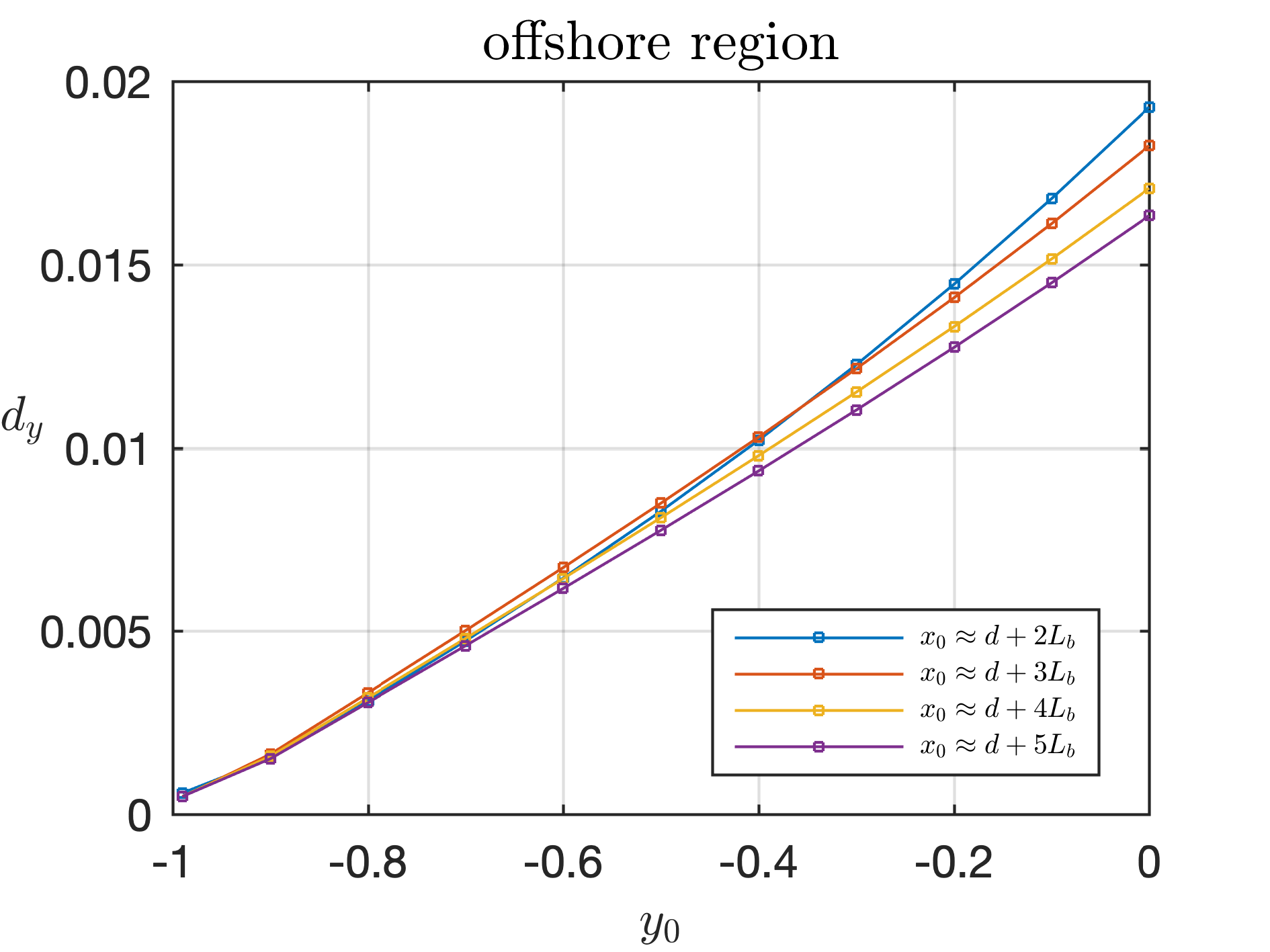}
        \caption{Horizontal ($d_x$) and vertical ($d_y$) particle displacements as functions of the initial depth ($y_0$) for different choices of $x_0$, considering $Fr = 0.125$. Recall that $L_b$ is the block length and $d$ is its distance traveled (see Figure \ref{schematic_figure_particles}). }
        
        \label{particle_displacement_low_Fr}
\end{figure}
\begin{figure}[!htb]
         \centering
         \includegraphics[scale=1]{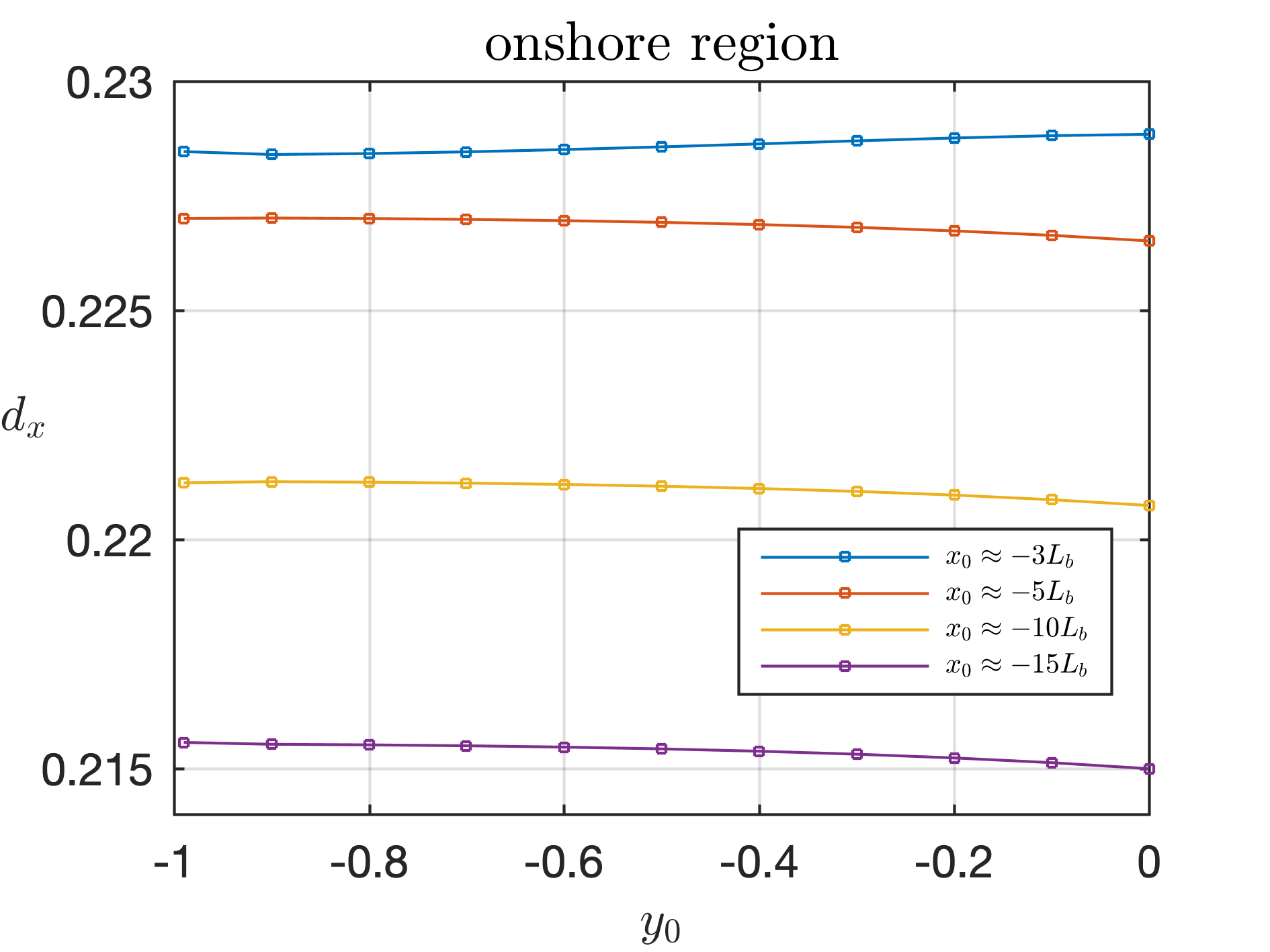}
         \includegraphics[scale=1]{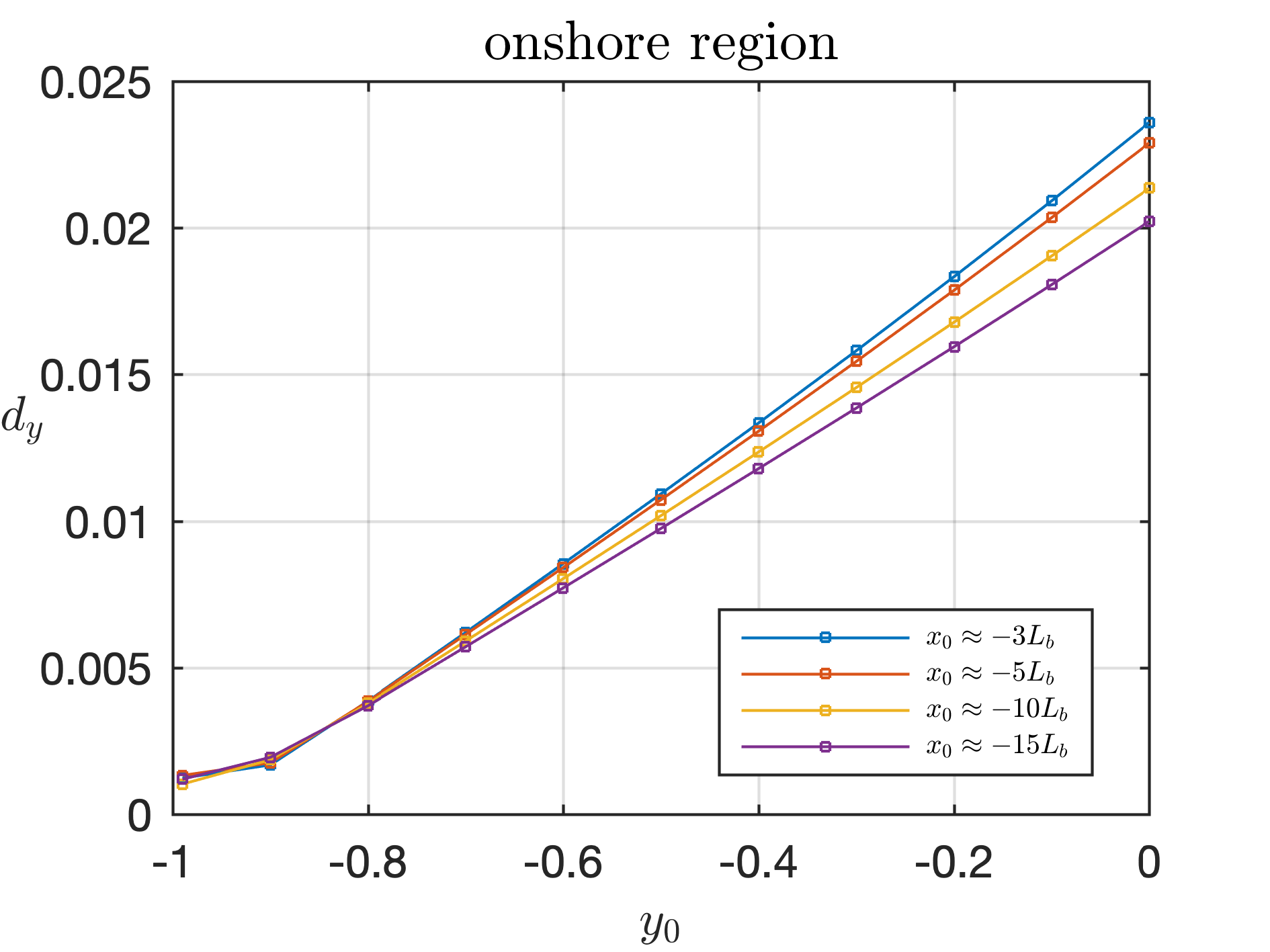}
         \includegraphics[scale=1]{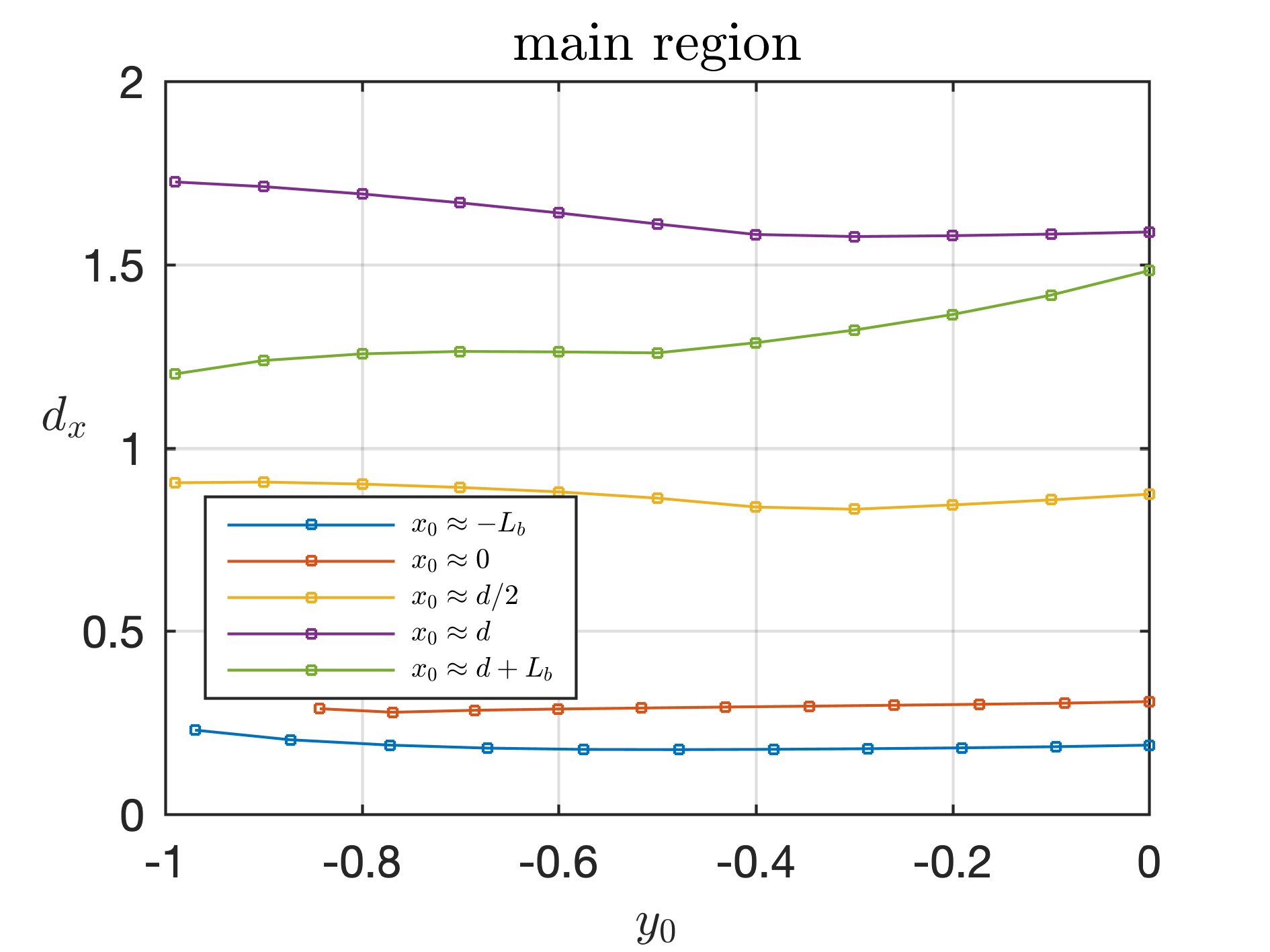}
         \includegraphics[scale=1]{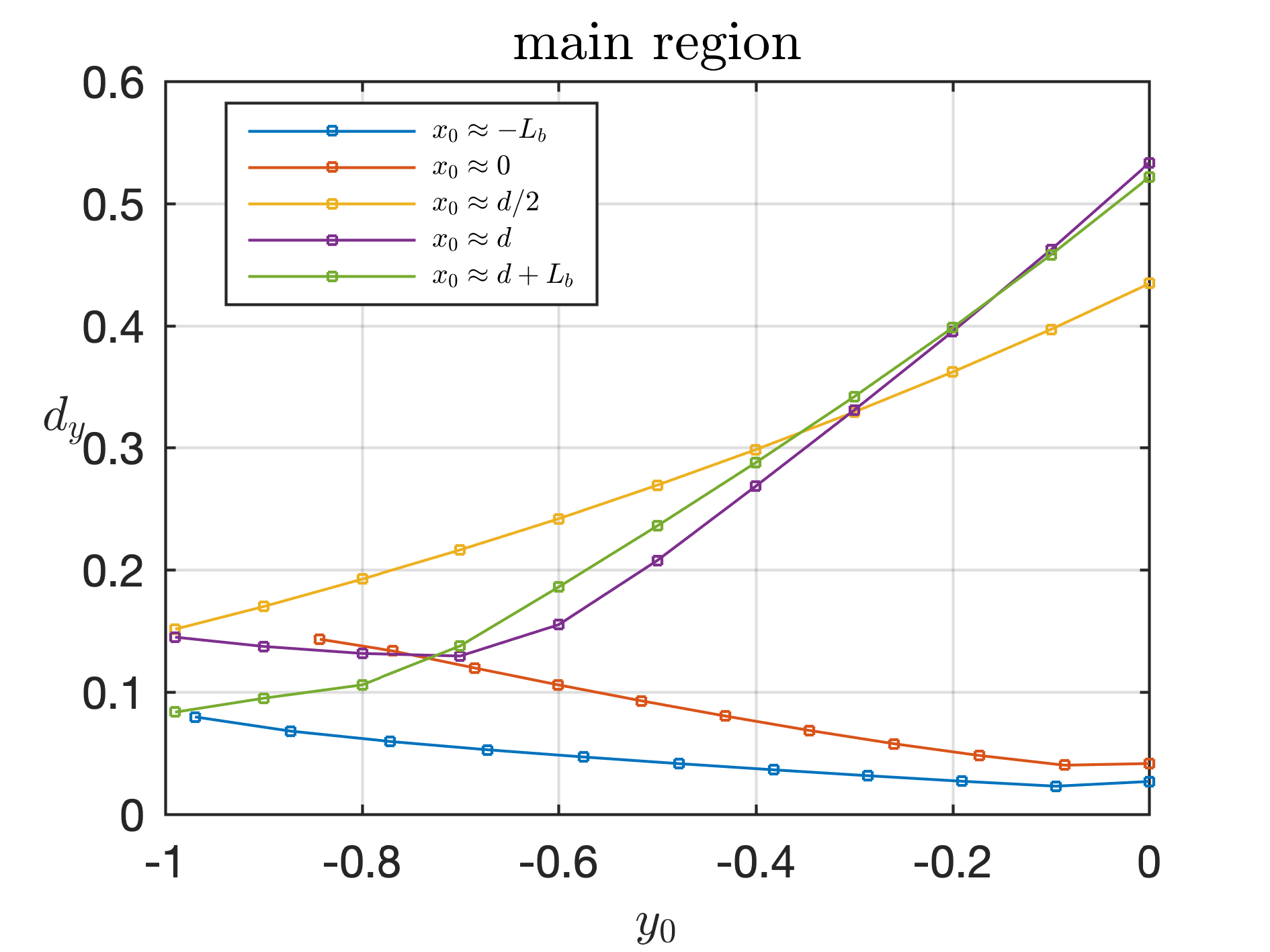}
         \includegraphics[scale=1]{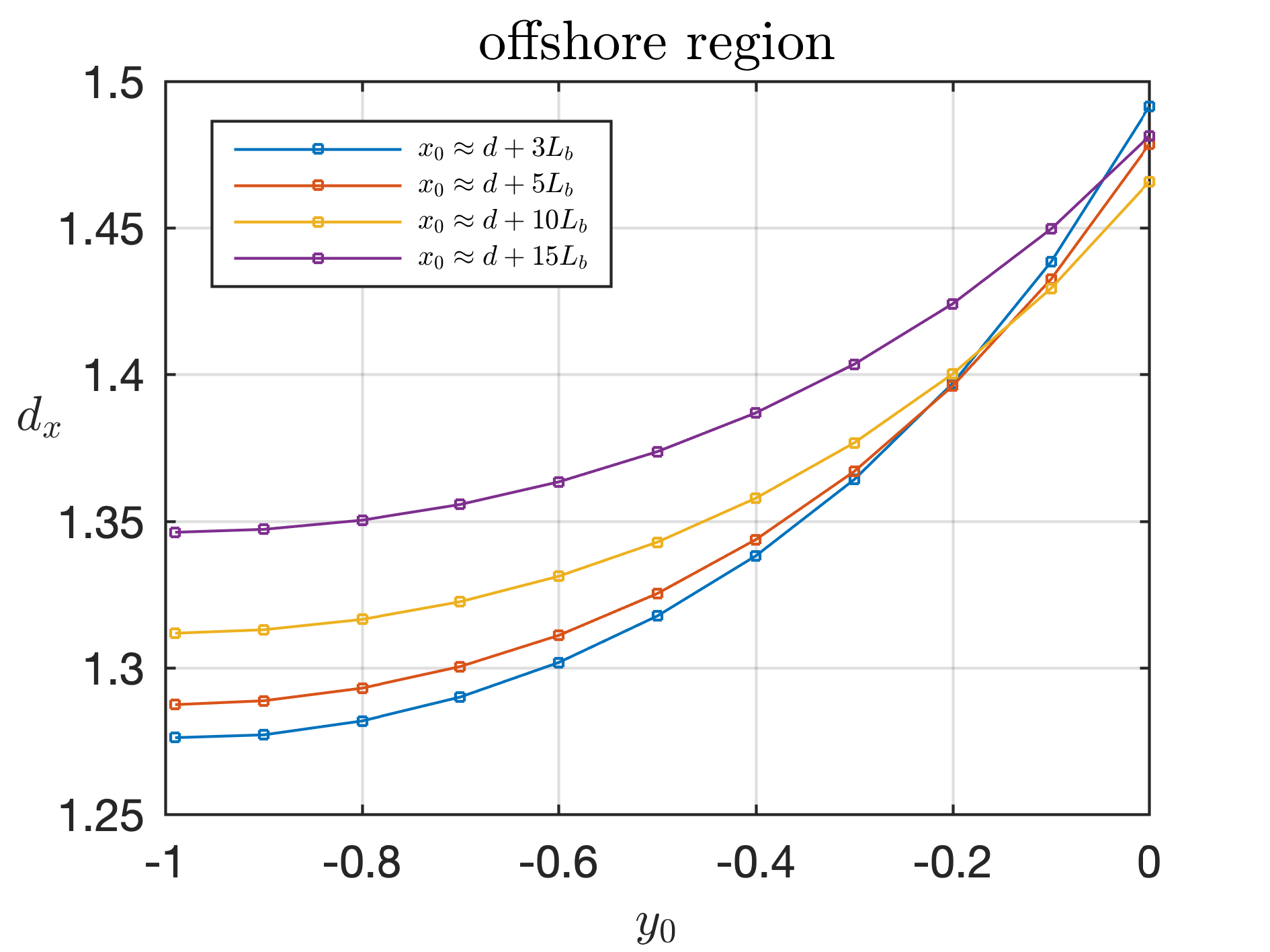}
         \includegraphics[scale=1]{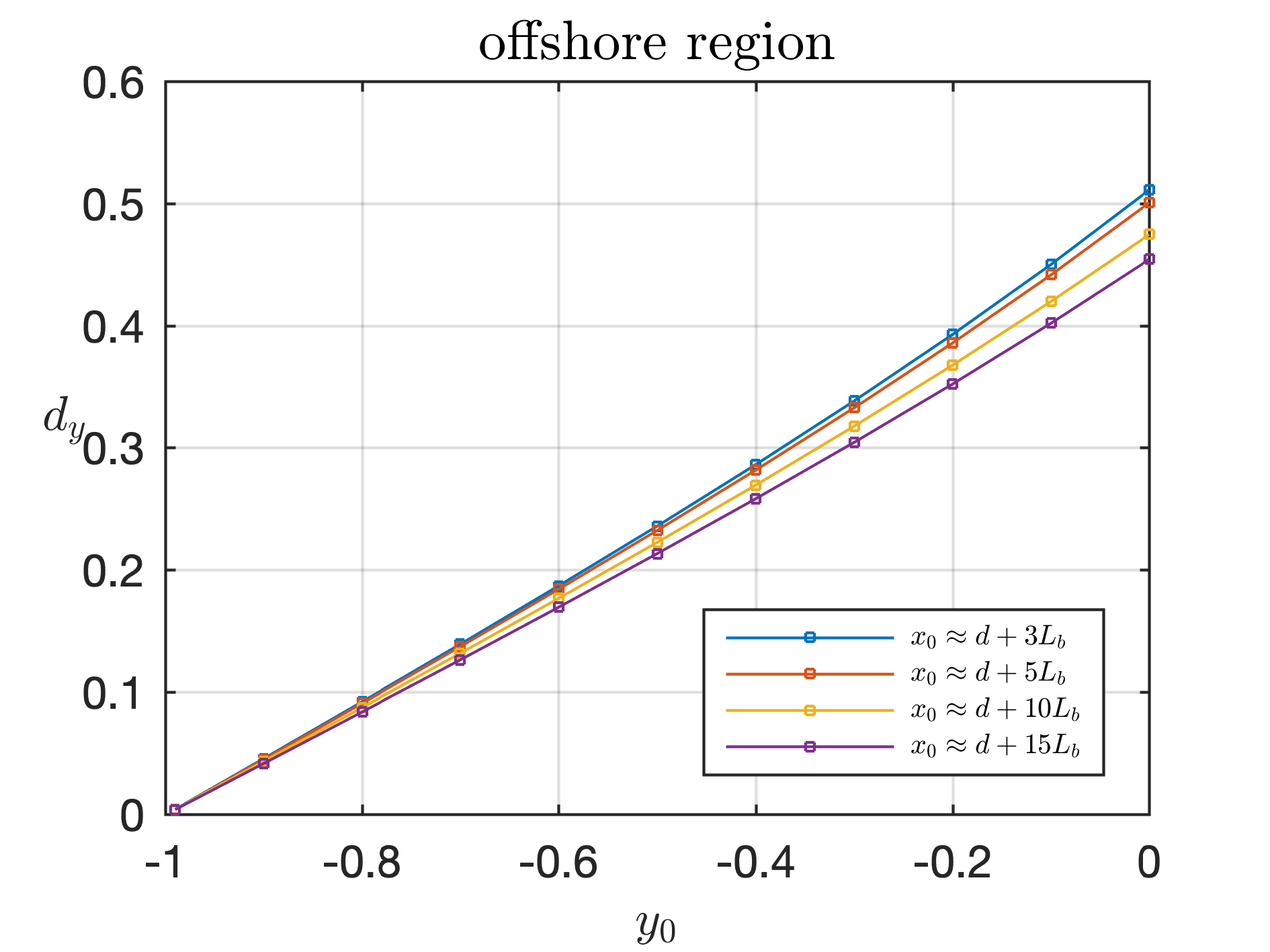}
        \caption{Horizontal ($d_x$) and vertical ($d_y$) particle displacements as functions of the initial depth ($y_0$) for different choices of $x_0$, considering $Fr = 1.2$. Recall that $L_b$ is the block length and $d$ is its distance traveled (see Figure \ref{schematic_figure_particles}). }
        \label{particle_displacement_high_Fr}
\end{figure}

Figures~\ref{particle_displacement_low_Fr} ($Fr = 0.125$) and \ref{particle_displacement_high_Fr} ($Fr = 1.2$) provide a systematic characterization of the horizontal and vertical particle displacements as functions of the initial depth $y_0$, for several initial horizontal positions $x_0$. In both cases, the largest excursions occur in the main region and are dominated by the horizontal component. For a given $x_0$, the horizontal displacement in this region varies only weakly with depth, indicating that the motion induced by the obstacle is transmitted across a substantial portion of the water column. In the onshore and offshore regions, the vertical displacement is small near the seabed and increases as the initial position approaches the free surface. 

Taken together, Figures~\ref{particle_displacement_low_Fr} ($Fr = 0.125$) and \ref{particle_displacement_high_Fr} ($Fr = 1.2$)  reveal a clear change in the mechanism governing particle motion as the Froude number increases. At \(Fr=0.125\), the particle displacement is predominantly associated with the motion of the seabed, with the horizontal displacement remaining relatively uniform across depth in the main region. At \(Fr=1.2\), the generated wave produces a much stronger vertical response, particularly for particles initially located near the free surface and away from the obstacle path. Thus, increasing the Froude number shifts the dominant contribution to particle motion from the direct forcing by the moving seabed toward the wave-induced flow. This transition is especially evident in the offshore region, where the bottom is effectively flat and the particle dynamics are governed solely by the propagating wave.



In the onshore and, more markedly, in the offshore region the vertical displacement almost vanishes near the bottom and grows approximately linearly as the initial depth decreases. However, the horizontal displacement does not exhibit this linear behavior with decreasing submergence, especially at a high Froude number in the offshore region. This is what one should expect, since these regions are never occupied by the obstacle and the bottom is effectively flat there, so that the particle response is governed by the propagating wave alone. The resulting profiles are similar to those reported by Nachbin and Ribeiro-Junior for periodic waves over a flat bottom \cite{Nachbin_RibeiroJr:2014}, which provides an independent consistency check on the trajectory solver in the region where the two problems become comparable.

\section{Conclusions} \label{Conclusion}

This work extends the framework introduced by the authors in \cite{Poletto:2025}, where conformal mapping was combined with spectral methods to solve the full Euler equations for waves generated by vertical seabed displacements, to the landslide problem, in which the obstacle translates horizontally with a prescribed, time-dependent velocity. Beyond the free surface, we computed the velocity field and the Lagrangian dynamics of fluid particles.  We also introduced a Hermite interpolation procedure that inserts arbitrary laboratory geometries into the solver directly.

The main methodological contribution is the trajectory system \eqref{system_particle_canonical}, together with the closed-form expressions \eqref{Gamma_equation} and \eqref{Lambda_equation} for the functions that carry the time dependence of the conformal map. Earlier Lagrangian computations built on Dyachenko-type maps were confined to travelling waves of permanent form, for which a change of frame renders the particle equations autonomous. That restriction is removed here: the scheme handles genuinely unsteady flows over moving topography at the same cost and the same spectral accuracy as the Eulerian solver, and it reproduces the kinematic boundary conditions to numerical precision, particles released on the free surface and on the seabed remaining there throughout the simulation. Nothing in the derivation is specific to landslides, so the same construction applies to other unsteady potential flows over a moving or deforming bottom.

The Lagrangian results provide a detailed picture of how particle motion is distributed throughout the fluid. Tracking particles in the onshore, main, and offshore regions shows that the largest excursions occur in the main region and are dominated by the horizontal component, whose magnitude varies only weakly with depth. The comparison between \(Fr=0.125\) and \(Fr=1.2\) further reveals a change in the dominant mechanism driving particle motion. At low Froude number, particle displacements are primarily associated with the direct motion of the seabed, whereas at high Froude number the generated wave produces a substantially stronger vertical response, particularly for particles initially located near the free surface and away from the obstacle path. In the offshore region, where the bottom is effectively flat, the particle dynamics are governed primarily by the propagating wave, and the displacement profiles approach those known for periodic waves over a flat bottom. These results illustrate that the Lagrangian formulation provides information on the spatial distribution and physical origin of particle motion that cannot be inferred from the free-surface evolution alone.

The model was validated against two independent sets of experiments by Whittaker et al. \cite{Whittaker:2015,Whittaker:2017}, covering Froude numbers from $0.125$ to $0.5$, with a quantitative error assessment reported in Table \ref{Norm_deviation}. 

On the physical side, we identified the terminal velocity of the slide (Froude number) as the parameter that controls the departure from linear theory. The amplitude and the acceleration of the bottom motion are secondary, and the shape of the obstacle is essentially immaterial once its area, height and centre of mass are fixed, a conclusion previously reached within linear and asymptotic theories and now confirmed within the full Euler equations for eight distinct geometries. As the Froude number approaches and exceeds unity, the nonlinear leading crest becomes substantially higher and faster than its linear counterpart, the onshore and offshore wave trains lose their symmetry, and the solution ceases to be invariant under $A \mapsto -A$, a symmetry that linear theory enforces by construction. These are precisely the quantities that matter for hazard assessment, which suggests that linear predictions should be treated with caution in the transcritical and supercritical regimes. We also found that at $Fr = 1.2$ the leading wave is largest for the smallest acceleration, reversing the trend documented at lower Froude numbers.

The velocity field displays the same asymmetry as the surface: at high Froude number the offshore region is far more energetic than the onshore one, whereas at $Fr = 0.125$ the field is nearly symmetric. Future work could consider more complex bottom forcings, including oscillatory or successive slides.


\subsection*{Declaration of generative AI use}

During the preparation of this work, the authors used Claude (Anthropic) exclusively for language editing and to improve the readability of the manuscript. The authors subsequently reviewed and edited the content as necessary and take full responsibility for the final content of the publication.

\begin{appendix}

\section{Derivation of the formulas for $\Gamma$ and $\Lambda$} \label{Gamma_Lambda_derivation}
This section is devoted to obtain the general formula for functions $\Lambda$ and $\Gamma$, shown in equations \eqref{Gamma_equation} and \eqref{Lambda_equation}.

Let us find a formula for $\Lambda$ before, by analyzing the boundary conditions. For the free surface, notice that 
\begin{equation*}
    \mathbf{Y}(\xi,t) = \zeta(\mathbf{X}(\xi,t),t).
\end{equation*}
From these relation, we obtain
\begin{equation*}
    \zeta_t(\mathbf{X}(\xi,t),t) =  \mathbf{Y}_t(\xi,t) - \dfrac{\mathbf{Y}_{\xi}(\xi,t)}{\mathbf{X}_{\xi}(\xi,t)}\mathbf{X}_t(\xi,t).
\end{equation*}

Let us omit the variables to ease the notation.

Now, we recall the kinematic equation and the formulas for the velocity potential, given by
\begin{align*}
    & {\zeta}_{t} + \phi_x{{\zeta}}_{x}-{{\phi}}_{y} = 0 \\
    & {\phi}_x  = \dfrac{1}{\mathbf{X}_{\xi}^2 + \mathbf{Y}_{\xi}^2}(\mathbf{\Phi}_{\xi}\mathbf{X}_{\xi} + \mathbf{\Psi}_{\xi}\mathbf{Y}_{\xi}) , \\
    & \phi_y = \dfrac{1}{\mathbf{X}_{\xi}^2 + \mathbf{Y}_{\xi}^2}(\mathbf{\Phi}_{\xi}\mathbf{Y}_{\xi} - \mathbf{\Psi}_{\xi}\mathbf{X}_{\xi}).
\end{align*}

After combining the four last equations we obtain
\begin{equation*}
    \mathbf{Y}_t - \dfrac{\mathbf{Y}_{\xi}}{\mathbf{X}_{\xi}}\mathbf{X}_t + \dfrac{1}{J}(\mathbf{\Phi}_{\xi}\mathbf{X}_{\xi} + \mathbf{\Psi}_{\xi}\mathbf{Y}_{\xi})\dfrac{\mathbf{Y}_{\xi}}{\mathbf{X}_{\xi}} = \dfrac{1}{J}(\mathbf{\Phi}_{\xi}\mathbf{Y}_{\xi} - \mathbf{\Psi}_{\xi}\mathbf{X}_{\xi}),
\end{equation*}
which can be simplified to
\begin{align*}
    & \mathbf{Y}_{t}\mathbf{X}_{\xi} - \mathbf{Y}_{\xi}\mathbf{X}_t = -\mathbf{\Psi}_{\xi}.
\end{align*}
Therefore, $\Lambda(\xi,0,t) = \dfrac{\mathbf{Y}_{t}\mathbf{X}_{\xi} - \mathbf{Y}_{\xi}\mathbf{X}_t}{J} = -\dfrac{\mathbf{\Psi}_{\xi}}{J}$.

Now we analyze the equations on the bottom boundary
\begin{equation*}
    Y(\xi,-D(t),t) = -1 + h(X(\xi,-D(t),t),t)
\end{equation*}
where $H(\xi,t) = h(X(\xi,-D(t),t),t)$ and $D(t) = 1 - \hat{H}(0,t) + \hat{Y}(0,t)$. Notice that
\begin{align*}
    & Y_t(\xi,-D(t),t) = H_t(\xi,t), \\
    & Y_{\xi}(\xi,-D(t),t) = H_{\xi}(\xi,t), \\
    & H_{\xi}(\xi,t)  = h_x(X(\xi,-D(t),t),t)X_{\xi}(\xi,-D(t),t), \\
    & H_t(\xi,t)  = h_x(X(\xi,-D(t),t),t)X_{t}(\xi,-D(t),t) + h_t(X(\xi,-D(t),t),t).
\end{align*}
Again, omitting the variables to ease the notation, leads to
\begin{equation*}
    \dfrac{Y_{bt}X_{b\xi} - X_{bt}Y_{\xi}}{J_b} = \dfrac{H_tX_{b\xi} - X_{bt}H_{\xi}}{J_b},
\end{equation*}
where the subindex $b$ refers to the functions evaluated at $\eta = -D(t)$ and $J_b = X_{b\xi}^2+Y_{b\xi}^2$. Combining the last equations leads to
\begin{align*}
    \dfrac{Y_{bt}X_{b\xi} - X_{bt}Y_{\xi}}{J_b} = \dfrac{h_tX_{b\xi}}{J_b}.
\end{align*}
Therefore, $\Lambda(\xi,-D(t),t) = \dfrac{h_tX_{b\xi}}{J_b}$.
Thus, since $\Gamma$ and $\Lambda$ are harmonic conjugated functions, we have the following system
\begin{equation}
\left\{
\begin{aligned}
     &\Lambda_{\xi\xi} + \Lambda_{\eta\eta} = 0 \quad \text{in} \quad \mathbb{C},\\
     &\Lambda(\xi,0,t) = -\frac{\mathbf{\Psi}_{\xi}(\xi,t)}{J} := f(\xi), \\
     &\Lambda(\xi,-D,t) = \frac{h_tx_{b\xi}}{J_b} := g(\xi).
 \end{aligned}
\right.
\end{equation}
 We now solve for $\Lambda$. Applying the Fourier transform to the system above gives
\begin{equation}
\left\{
\begin{aligned}
     &-k^2\widehat{\Lambda} + \widehat{\Lambda}_{\eta\eta} = 0, \label{edp_FLambda}\\
     &\widehat{\Lambda}(k,0,t) = \widehat{f}, \\
     &\widehat{\Lambda}(k,-D,t) = \widehat{g}.
 \end{aligned}
\right.
\end{equation}
 First, for $k \neq 0$, the solution for \eqref{edp_FLambda} is
 \begin{equation*}
     \widehat{\Lambda}(k,\eta,t) = A(k,t) e^{k \eta} + B(k,t) e^{-k\eta},
 \end{equation*}
 which evaluated at $\eta = 0$ and $\eta = -D$ gives
 \begin{align*}
     &A(k,t) + B(k,t) = \widehat{f} \\
     & A(k,t)e^{-kD} + B(k,t)e^{kD} = \widehat{g}. 
 \end{align*}
 Isolating $A(k,t)$ above and substituting it results in
 \begin{equation*}
     (\widehat{f} - B(k,t))e^{-kD} + B(k,t) e^{kD} = \widehat{g}
 \end{equation*}
Hence, we can obtain 
\begin{align*}
    & B(k,t)(e^{kD} - e^{-kD}) = \widehat{g} - \widehat{f}e^{-kD} \\
    & B(k,t) = \frac{\widehat{g} - \widehat{f}e^{-kD}}{2\sinh{(kD)}}.
\end{align*}
Thus
\begin{align*}
    A(k,t) & = \widehat{f} - \frac{\widehat{g} - \widehat{f}e^{-kD}}{2\sinh{(kD)}} \\
    & = \frac{\widehat{f}(e^{kD} - e^{-kD}) - \widehat{g} + \widehat{f}e^{-kD}}{2\sinh{(kD)}} \\
    & = \frac{\widehat{f}e^{kD} - \widehat{g}}{2\sinh{(kD)}}
\end{align*}
Hence, the expression for $\widehat{\Lambda}(k\neq 0,\eta,t)$ is
\begin{align*}
    \widehat{\Lambda}(k,\eta,t) & = \frac{\widehat{f}e^{kD} - \widehat{g}}{2\sinh{(kD)}}e^{k\eta} + \frac{\widehat{g} - \widehat{f}e^{-kD}}{2\sinh{(kD)}}e^{-k\eta} \\
    & = \frac{1}{2\sinh{(kD)}}\left[\widehat{f}(e^{k(D + \eta)} - e^{-k(D + \eta)}) - \widehat{g}(e^{k\eta} - e^{-k\eta}) \right] \\
    & = \frac{\sinh{(k(D + \eta))}}{\sinh{(kD)}}\widehat{f}(k,t) - \frac{\sinh{(k\eta)}}{\sinh{(kD)}}\widehat{g}(k,t)
\end{align*}
Considering the case where $k \rightarrow 0$ we have
\begin{equation*}
    \Lambda(\xi,\eta,t) = \mathcal{F}^{-1}_{k \neq 0}\left[\frac{\sinh{(k(D + \eta))}}{\sinh{(kD)}}\widehat{f}(k,t) - \frac{\sinh{(k\eta)}}{\sinh{(kD)}}\widehat{g}(k,t) \right] + \frac{(D + \eta)}{D}\widehat{f}(0,t) - \frac{\eta}{D}\widehat{g}(0,t)
\end{equation*}

Through the Cauchy Riemann equation, we get
\begin{equation}\label{gamma_xi}
    \Gamma_{\xi} = \Lambda_{\eta} = \mathcal{F}^{-1}_{k \neq 0}\left[\frac{k\cosh{(k(D + \eta))}}{\sinh{(kD)}}\widehat{f}(k,t) - k\frac{\cosh{(k\eta)}}{\sinh{(kD)}}\widehat{g}(k,t) \right] + \frac{\widehat{f}(0,t)}{D} - \frac{\widehat{g}(0,t)}{D}
\end{equation}
Applying the Fourier transform
\begin{equation*}
    ik\widehat{\Gamma}(k\neq 0,\eta,t) = \frac{k\cosh{(k(D + \eta))}}{\sinh{(kD)}}\widehat{f}(k,t) - k\frac{\cosh{(k\eta)}}{\sinh{(kD)}}\widehat{g}(k,t),
\end{equation*}
which, together with equation \eqref{gamma_xi}, gives
\begin{align*}
    \Gamma(\xi,\eta,t) & = \mathcal{F}^{-1}_{k \neq 0}\left[\frac{-i\cosh{(k(D + \eta))}}{\sinh{(kD)}}\widehat{f}(k,t) + i\frac{\cosh{(k\eta)}}{\sinh{(kD)}}\widehat{g}(k,t) \right] + \left(\frac{\widehat{f}(0,t)}{D} - \frac{\widehat{g}(0,t)}{D} \right)\xi.
\end{align*}

Evaluating at the free surface, the formula for $\Gamma(\xi,\eta = 0,t)$ reads
\begin{equation*}
    \Gamma(\xi,0,t) = \mathcal{F}^{-1}_{k \neq 0} \left [ i\coth{(kD)}\left(\frac{1}{\cosh{(kD)}}\mathcal{F}\left[\frac{h_tX_{b\xi}}{J_b} \right] + \mathcal{F}\left[\frac{\mathbf{\Psi}_{\xi}(\xi,t)}{J}\right]\right) \right] + \widehat{\Gamma}(0,0,t)
\end{equation*}

Now, we can define $\Gamma_0(\xi,t) := \Gamma(\xi,0,t)$ and 
\begin{equation*}
    \widehat{P}(k,t) = \frac{1}{\cosh{(kD)}}\mathcal{F}\left[\frac{h_tX_{b\xi}}{J_b} \right] + \mathcal{F}\left[\frac{\mathbf{\Psi}_{\xi}(\xi,t)}{J}\right].
\end{equation*}
Hence, we can write
\begin{equation*}
    \Gamma_0(\xi,t) = \mathcal{F}^{-1}\left[i\coth{(kD)}\widehat{P}(k) \right]
\end{equation*}

In order to compute $\widehat{\Gamma}_0(0,t)$ we must consider that there exists an analytical function given by
\begin{equation*}
    M_{\xi}(\xi,t) = P(\xi,t),
\end{equation*}
which allows us to evaluate
\begin{equation*}
    \lim_{k\rightarrow 0} i \coth{(kD)}\widehat{M}_{\xi}(k,t) = -\frac{\left<M(\xi,t) \right>}{D}.
\end{equation*}
Therefore, the operator $\mathcal{C}$ given in equation \eqref{operator_C} is also computed by
\begin{equation*}
    \mathcal{C}\left[P\right] = \mathcal{C}_0[P] - \left< \mathbf{X}_{\xi}\mathcal{C}_0\left[P\right] 
 + \mathbf{Y}_{\xi}\frac{\mathbf{\Psi}_{\xi}}{J}\right>.
\end{equation*}

\end{appendix}


\section*{Acknowledgments}
This study was financed in part by the Coordena\c{c}\~{a}o de Aperfei\c{c}oamento de Pessoal de N\'{\i}vel Superior --- Brasil (CAPES) --- Finance Code 001. The work of M.V.F.  was funded by the Vicerrectorado de
Investigaci{\' o}n (VRI) at the PUCP through grant DFI-2026-PI1434.


\end{document}